\documentclass[aps,prl,twocolumn,superscriptaddress]{revtex4-2}

\usepackage{amssymb}
\usepackage{natbib}
\usepackage{multirow}
\usepackage{bm}
\usepackage{amsmath}
\usepackage{graphicx}
\usepackage{epstopdf}
\usepackage{subfigure}
\usepackage{natbib}
\usepackage{epsfig}
\usepackage{amsfonts}
\usepackage{mathrsfs}
\usepackage{braket}
\usepackage{soul}
\usepackage{tabularx}
\usepackage{xcolor}
\usepackage[toc,page,title,titletoc,header]{appendix}

\usepackage{dsfont,amsthm,amsbsy}

\usepackage{lineno}
\begin{document}

\title{Anomalous Landau level gaps near magnetic transitions in monolayer WSe$_2$}

\author{Benjamin A. Foutty}
\affiliation{Geballe Laboratory for Advanced Materials, Stanford, CA 94305, USA}
\affiliation{Department of Physics, Stanford University, Stanford, CA 94305, USA}

\author{Vladimir Calvera}
\affiliation{Department of Physics, Stanford University, Stanford, CA 94305, USA}

\author{Zhaoyu Han}
\affiliation{Department of Physics, Stanford University, Stanford, CA 94305, USA}

\author{Carlos R. Kometter}
\affiliation{Geballe Laboratory for Advanced Materials, Stanford, CA 94305, USA}
\affiliation{Department of Physics, Stanford University, Stanford, CA 94305, USA}

\author{Song Liu}
\affiliation{Department of Mechanical Engineering, Columbia University, New York, NY, 10027, USA}

\author{Kenji Watanabe}
\affiliation{Research Center for Electronic and Optical Materials, National Institute for Materials Science, 1-1 Namiki, Tsukuba 305-0044, Japan}

\author{Takashi Taniguchi}
\affiliation{Research Center for Materials Nanoarchitectonics, National Institute for Materials Science, 1-1 Namiki, Tsukuba 305-0044, Japan}

\author{James C. Hone}
\affiliation{Department of Mechanical Engineering, Columbia University, New York, NY, 10027, USA}

\author{Steven A. Kivelson}
\affiliation{Department of Physics, Stanford University, Stanford, CA 94305, USA}

\author{Benjamin E. Feldman}
\email{bef@stanford.edu}
\affiliation{Geballe Laboratory for Advanced Materials, Stanford, CA 94305, USA}
\affiliation{Department of Physics, Stanford University, Stanford, CA 94305, USA}
\affiliation{Stanford Institute for Materials and Energy Sciences, SLAC National Accelerator Laboratory, Menlo Park, CA 94025, USA}

\def\EZb{E_Z^0}

\begin{abstract}    
    First-order phase transitions produce abrupt changes to the character of both ground and excited electronic states. Here we conduct electronic compressibility measurements to map the spin phase diagram and Landau level (LL) energies of monolayer WSe$_2$ in a magnetic field. We resolve a sequence of first-order phase transitions between completely spin-polarized LLs and states with LLs of both spins. Unexpectedly, the LL gaps are roughly constant over a wide range of magnetic fields below the transitions, which we show reflects a preference for opposite spin excitations of the spin-polarized ground state. These transitions also extend into compressible regimes, with a sawtooth boundary between full and partial spin polarization. We link these observations to the important influence of LL filling on the exchange energy beyond a smooth density-dependent contribution. Our results show that WSe$_2$ realizes a unique hierarchy of energy scales where such effects induce re-entrant magnetic phase transitions tuned by density and magnetic field.
\end{abstract}

\maketitle
\section{Introduction} 
Electronic systems with degeneracies arising from internal quantum degrees of freedom are often susceptible to forming ordered ground states driven by many-body interactions. The quantum Hall regime, in which Landau levels (LLs) effectively quench kinetic energy, provides a model platform to study such  
phases and the transitions between them. In particular, the relative energies of LLs with distinct spin and/or valley indices can often be modified by experimental tuning knobs which affect both the many-body ground and excited states in these systems \cite{sondhi_skyrmions_1993,piazza_first-order_1999,tutuc_-plane_2001,gunawan_valley_2006,nomura_quantum_2006,kott_valley-degenerate_2014,hunt_direct_2017,shi_bilayer_2022,shi_odd-_2020,maryenko_polarization-dependent_2014,falson_review_2018}. However, the nature of charge excitations near transitions between competitive LLs depends sensitively on details of the LL energetics and can be difficult to directly probe.

\begin{figure*}[t!]
    \renewcommand{\thefigure}{\arabic{figure}}
    \centering
    \includegraphics[scale =1.0]{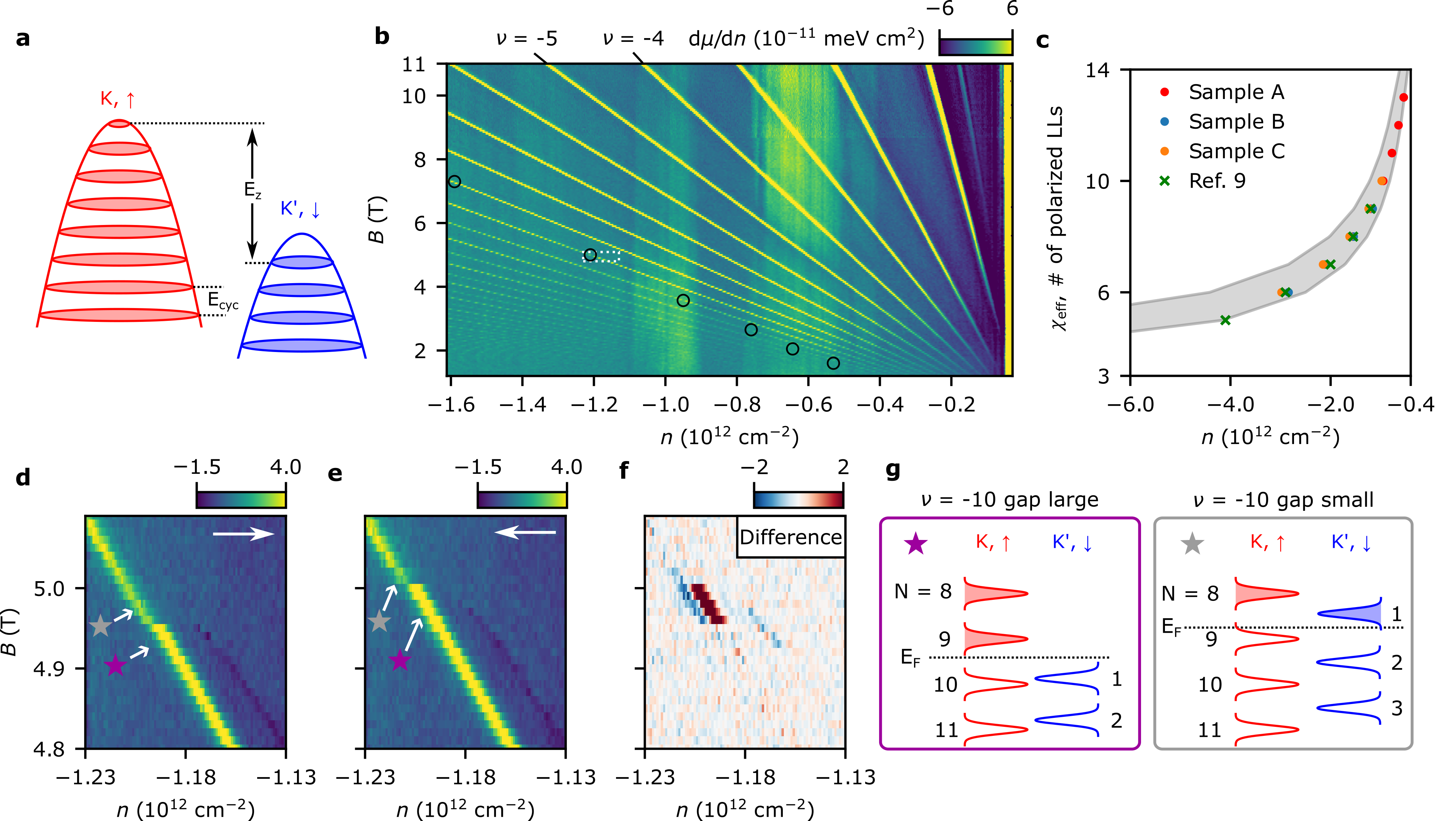}
    \caption{\textbf{First-order Landau level (LL) phase transitions in monolayer WSe$_2$.} \textbf{a}, Schematic of the valence band LL structure in monolayer WSe$_2$. The two relevant bands are spin up at valley $K$ and spin down at valley $K^\prime$, which are split by a density-dependent effective Zeeman energy $E_Z$. $E_{\rm{cyc}}$ is the cyclotron energy. \textbf{b}, Inverse electronic compressibility d$\mu$/d$n$ in monolayer WSe$_2$ as a function of hole density $n$ and perpendicular magnetic field $B$. Black circles mark sharp drops in the magnitude of d$\mu$/d$n$ along incompressible LL gaps. Broad vertical features are artifacts due to long a.c. charging time of the sample (Methods). \textbf{c}, Spin susceptibility $\chi_{\textrm{eff}}$ as a function of $n$, determined from the densities of the LL transitions (such as those highlighted in \textbf{b}) from three distinct samples. We also present data from Ref. \cite{shi_odd-_2020} and theoretical predictions based on quantum Monte Carlo calculations shaded in gray for comparison. \textbf{d-e}, Zoom-ins of d$\mu$/d$n$ in the white box in panel \textbf{b}, with the density swept in opposite directions (large white arrows). \textbf{f}, Difference between panels $\textbf{d-e}$, demonstrating pronounced hysteresis from a first-order phase transition. \textbf{g}, Schematics of LL energies that respectively correspond to the starred positions in \textbf{d-e}; note that due to the hole carriers, states are filled from the top downward. $N$ denotes the LL orbital index and $E_{\rm{F}}$ is the Fermi level.}
    \label{fig:Fig1}
\end{figure*}

Monolayer semiconducting transition metal dichalcogenides (TMDs) realize a distinctive LL structure due to their hierarchy of energy scales. Strong spin-orbit coupling near the valence band maxima at valleys $K$ and $K^\prime$ causes the relevant low-energy bands to be spin-valley locked, so that only a single Ising spin orientation is relevant at each valley \cite{xu_spin_2014}. This degree of freedom forms a generalized isospin, which we refer to as spin in the rest of the text.
A combination of the large effective mass and the additive contributions of orbital and Berry-curvature effects produces a 
single particle Zeeman splitting of the valence band $\EZb$ that is large relative to the cyclotron energy $E_{\rm{cyc}}$ in monolayer WSe$_2$ ($\EZb / E_{\rm{cyc}} \approx 2)$ \cite{gustafsson_ambipolar_2018}. The large effective mass also enhances the relative importance of interactions,
such that the dimensionless parameter $r_s$ is of the order of $5-10$ at achievable carrier densities ~\cite{wang_valley-_2017}. 

Prior work has shown that these interactions drive a density-dependent exchange enhancement of the effective Zeeman energy $E_Z$ \cite{gustafsson_ambipolar_2018,shi_odd-_2020,movva_density-dependent_2017}. This increases the spin-splitting of the valence bands as the hole density decreases and leads to preferential occupation of fully spin-polarized LLs at low densities (Fig. \ref{fig:Fig1}a) \cite{wang_valley-_2017}. At higher densities, both `majority' and `minority' spin Zeeman-split LLs are occupied, causing alternating LL gap sizes dominated by even or odd integers \cite{movva_density-dependent_2017,larentis_large_2018,pisoni_interactions_2018,gustafsson_ambipolar_2018,shi_odd-_2020}. Recent studies have noted the possibility of first-order phase transitions at the crossover between these limits in monolayer WSe$_2$ and related systems, but hysteresis has not been observed and a detailed understanding of the LL energetics as the system transitions from fully to partially spin-polarized has until now been lacking \cite{shi_odd-_2020,li_spontaneous_2020,shih_spin-selective_2023}.

\begin{figure*}[t!]
    \renewcommand{\thefigure}{\arabic{figure}}
    \centering
    \includegraphics[scale =1.0]{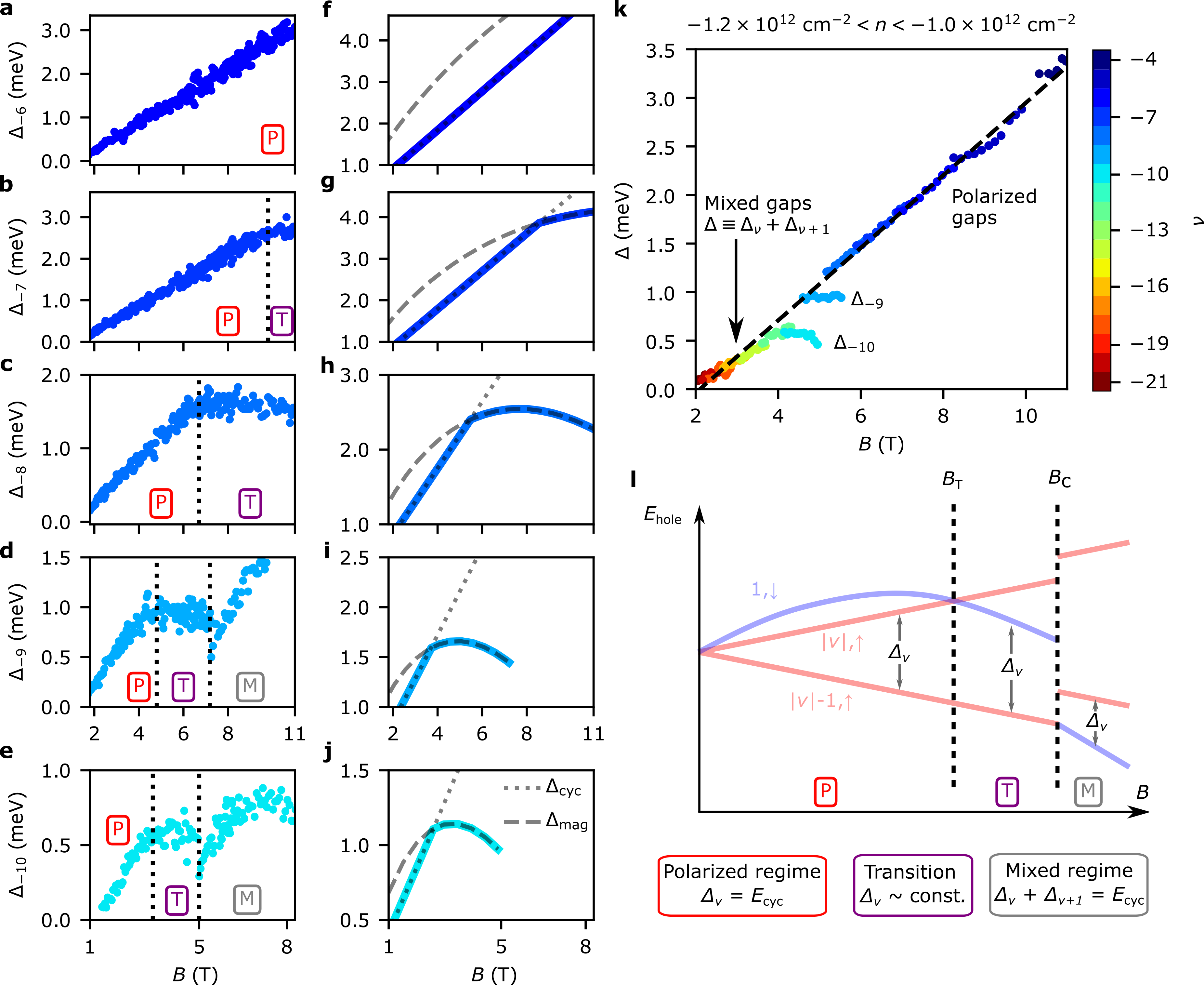}
    \caption{\textbf{Magnetic field dependence of LL gaps.} \textbf{a-e}, Experimentally measured LL gaps $\Delta_{\nu}$ as a function of $B$  for fixed integer values of the filling factor, $\nu$. We observe three distinct behaviors: fully polarized LLs with $\Delta_\nu = E_{\rm{cyc}}$ (labeled by `P'), a transition region where $\Delta_\nu$ plateaus (labeled by `T'), and a `mixed' regime where both spins are occupied and $\Delta_{\nu} + \Delta_{\nu+1} = E_{\rm{cyc}}$ (labeled by `M'). Vertical dotted lines indicate boundaries separating these behaviors. \textbf{f-j}, Corresponding theoretically predicted LL gaps from Hartree-Fock calculations with RPA-screened interactions. We show $\Delta_{\rm{cyc}}$ (dotted lines), the gap to the next unoccupied majority spin LL, and $\Delta_{\rm{sf}}$ (dashed lines), the gap to the $N = 1$ minority spin LL. The LL gap is the smaller of these (thick colored lines). \textbf{k}, Measured gaps $\Delta$ as a function of $B$ in the range $-1.2\times 10^{12}$ cm$^{-2} < n < -1.0\times 10^{12}$ cm$^{-2}$. Color indicates the filling factors of different gaps: polarized and `transition' gaps (from $\nu = -4$ to $\nu = -10$) are plotted individually ($\Delta \equiv \Delta_{\nu}$), while the pairwise sum of gaps are plotted for even $\nu \le -11$. The dashed black line is a linear fit (excluding the `transition' gaps). \textbf{l}, Cartoon of the LL energies relevant to $\Delta_{\nu}$, given in terms of $E_{\rm{hole}} = -E$, so LLs are filled from the bottom up. At $B_{\rm{T}}$, the lowest-energy LL above the gap switches from majority to minority spin, while at $B_{\rm{C}}$, the system undergoes a first-order phase transition to the mixed regime.
    }
    
    \label{fig:Fig2}
\end{figure*}

In this work, we use a scanning single-electron transistor (SET) to measure the inverse electronic compressibility, d$\mu$/d$n$, of valence band holes in monolayer WSe$_2$ in a perpendicular magnetic field. At the crossover between fully spin-polarized LLs and the lowest-energy minority spin LL being filled, we resolve first-order phase transitions, including hysteresis in the LL gaps and adjoining sharp tails of negative compressibility that extend outward into nearby compressible electronic states. Surprisingly, the LL gaps are roughly constant over a wide range of magnetic fields below these phase transitions. Through high-resolution measurements of the thermodynamic LL gaps and the first-order phase transitions, together with suppporting theoretical calculations, we systematically characterize the nature of low-energy charge excitations throughout the phase diagram. Collectively, these indicate multiple reorderings of the LL structure as we vary carrier density and magnetic field. Our results provide a straightforward way, in the correlation-dominated regime, to understand the spin character and energies of ground and excited electronic states.

\begin{figure*}[t!]
    \renewcommand{\thefigure}{\arabic{figure}}
    \centering
    \includegraphics[scale =1.0]{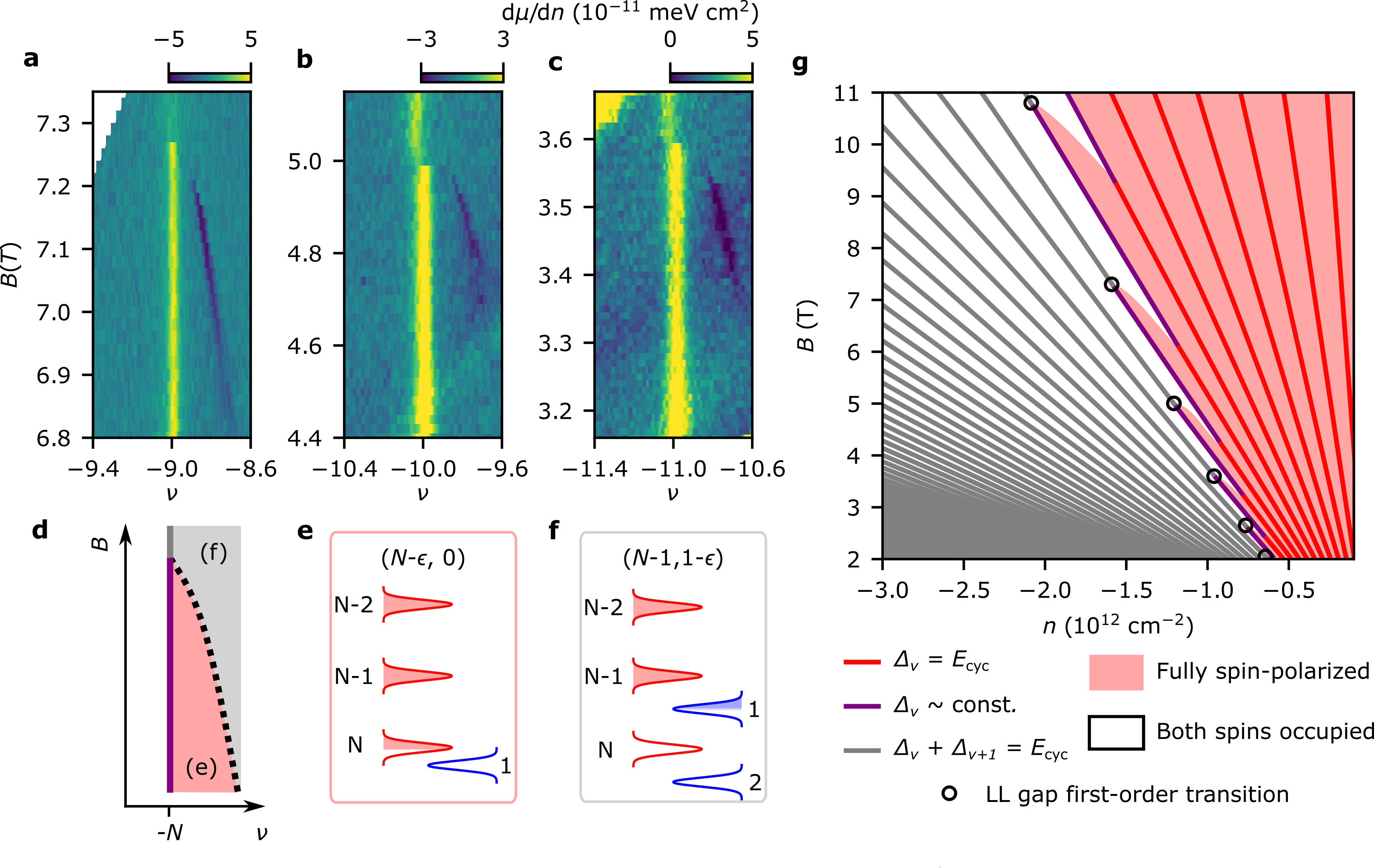}
    \caption{\textbf{Spin transitions within partially filled LLs and full phase diagram.} \textbf{a-c}, High-resolution measurements of d$\mu$/d$n$ near phase transitions at $\nu = -9, -10,-11$, highlighting the negative compressibility which extends into the adjacent LL. Panel $\textbf{a}$ is measured at temperature $T = 1.6$ K, while $\textbf{b-c}$ are at $T = 0.35$ K. The small shift in $\nu$ of the incompressible peak that occurs across the phase transition in \textbf{b-c} likely reflects asymmetry in the broadening of the crossing LLs (Supplementary Sec. 9). \textbf{d}, Schematic showing regions of distinct partially occupied LLs as a function of $B$ when $|\nu| \le N$ for a given integer $N$. The pink shaded region corresponds to a fully-polarized phase, with filling $(|\nu_\uparrow|,|\nu_\downarrow|) = (N-\epsilon,0)$, where $\nu_{\uparrow(\downarrow)}$ is the filling factor of holes in the spin-$\uparrow(\downarrow)$ sector and $0<\epsilon<1$. The light gray region is a mixed state with $(|\nu_\uparrow|,|\nu_\downarrow|) = (N-1,1-\epsilon)$, i.e. a minority spin LL is partially filled. \textbf{e-f}, Schematic of the LL orderings in each compressible phase, with red spin majority and blue spin minority LLs labelled by their respective orbital indices. \textbf{g}, Schematic depiction of the full spin phase diagram as a function of $n$ and $B$. Lines indicate the LL gap behavior, while shading indicates the spin character of the filled states. }
    \label{fig:Fig3}
\end{figure*}

\section{First-order spin transitions} 

In Fig. \ref{fig:Fig1}b, we present a Landau fan of d$\mu/$d$n$ as a function of carrier density $n$ and perpendicular magnetic field $B$. Across data from three distinct devices, we observe qualitatively similar behavior, though we focus on data from a single sample through most of the main text (see Supplementary Sec. 1 for a detailed comparison). Most of the incompressible features can be identified via their slope in the $n-B$ plane as integer quantum Hall gaps, which occur at all integer filling factors $\nu$. Additionally, we note fractional quantum Hall states in the lowest LL consistent with previous reports \cite{shi_odd-_2020,pack_charge-transfer_2023} (Supplementary Sec. 2).

Following constant integer $\nu$, we observe abrupt drops in the magnitude of the LL gaps (marked by black circles in Fig. \ref{fig:Fig1}b) that occur when the first minority spin LL crosses the highest occupied majority spin LL, corresponding to a transition from full to partial spin polarization \cite{shi_odd-_2020}. We determine the effective spin susceptibility $\chi_{\textrm{eff}}$ by identifying the number of polarized LLs at the phase transitions \cite{shi_odd-_2020,vakili_spin_2004,maryenko_polarization-dependent_2014}. Note this is actually a response to a finite field, and can strictly be identified as the ``susceptibility'' in the noninteracting limit. We find that $\chi_{\textrm{eff}}$ extends up to 13, corresponding to a $g$-factor of roughly $35$ at the lowest density transition that we can resolve (Fig. \ref{fig:Fig1}c). The observed increase of $\chi_{\textrm{eff}}$ with decreasing hole density is broadly consistent with previous quantum Monte Carlo calculations of a 2D electron gas, shaded in gray \cite{vakili_spin_2004,attaccalite_correlation_2002} (Supplementary Sec. 8).

We resolve hysteresis in both the LL gaps and adjacent negative compressibility features (discussed in detail below) as shown in Fig. \ref{fig:Fig1}d-f, direct evidence of first-order phase transitions. As the hole density is swept in opposite directions, the sharp drop in gap size occurs at distinct magnetic fields (hysteresis is also evident upon sweeping $B$, see Supplementary Sec. 1). The behavior reflects spontaneous polarization switching of the last occupied LL (Fig. \ref{fig:Fig1}g) \cite{li_spontaneous_2020,shih_spin-selective_2023}. Specifically, the large gap is stabilized when sweeping from a fully polarized phase, where exchange interactions favor maintaining maximum spin polarization and enhance the effective $g$-factor \cite{ando_theory_1974}. We only resolve hysteresis as the first (orbital index $N = 1$) minority spin LL becomes competitive with the valence majority spin level; at lower magnetic fields at the same density where LLs with higher indices cross, there are no sharp changes in the measured gaps (Supplementary Sec. 3). This indicates that the additional $g$-factor enhancement is suppressed when both majority and minority spin LLs are occupied, destroying the expected first-order transition or rendering it undetectably weak.

\section{Anomalous LL gap scaling}

Our measurements encode information not only about changes in the occupied LLs, but also about excited states immediately above the Fermi level. To study the LL energetics in the vicinity of the spin phase transitions, we integrate d$\mu$/d$n$ to obtain the thermodynamic gaps $\Delta_{\nu}$ at integer filling factors $\nu$ (Fig. \ref{fig:Fig2}). For each integer quantum Hall gap, we observe three distinct behaviors as a function of magnetic field. We relate these behaviors to distinct ground states and their lowest-energy charge excitations, as detailed below.

At fixed filling factor and low fields, highlighted by a red `P' in Fig. \ref{fig:Fig2}a-e, the LLs are fully spin polarized. We observe a linear field dependence of these gaps, indicating that they are set by the cyclotron energy $E_{\rm{cyc}} = \frac{\hbar e B}{m^*}$ to the next spin majority LL. We extract an effective mass $m^*\approx 0.31m_e$ from the linear slope, where $\hbar$ is the reduced Planck's constant and $e$ and $m_e$ are the electron charge and mass ($m^*$ depends weakly on sample, see Supplementary Sec. 1). At fixed filling factor and sufficiently high field, highlighted by a gray `M' in Fig. \ref{fig:Fig2}, the LLs are in a `mixed' regime where both spins are occupied. Individual gaps grow and shrink as the density-dependent effective Zeeman energy changes the relative spacing between LLs of different spin, but the pairwise sum of gaps  $\Delta_{\nu} + \Delta_{\nu+1}$ at a given density matches the cyclotron energy with a similar effective mass to that of the polarized LLs (Fig.~\ref{fig:Fig2}k, Supplementary Sec. 4). Both the polarized and mixed regimes can be well-described by the previously considered model in which LLs are affected by a smooth density-dependent Zeeman enhancement but are otherwise unchanged energetically \cite{gustafsson_ambipolar_2018,shi_odd-_2020}. Our experiments, however, demonstrate a more complicated behavior at the transition between these regions.

Between the polarized and mixed regimes, we observe that each LL gap plateaus as the field is increased preceding its first-order phase transition. Remarkably, the range of magnetic fields over which the gaps are flat, highlighted by a purple `T' in Fig. \ref{fig:Fig2}, can extend over several Tesla (e.g. between 6.5 and 11 T for $\nu = -8$). At a given density, two gaps (the final two LL gaps that are not in the mixed regime) are approximately field independent and diverge from the cyclotron energy scale. This is best illustrated by plotting the LL gaps within a fixed density range (Fig.~\ref{fig:Fig2}k). It is surprising that for any given hole density, multiple LL gaps are set by a scale comparable to, but smaller than the cyclotron energy. Similar plateaus persist across all three samples which were fabricated independently and exfoliated from bulk WSe$_2$ crystals from different sources, as well as a fourth device of Bernal bilayer WSe$_2$ (Supplementary Secs. 1, 4, and 5). This consistency indicates intrinsic and generic behavior unrelated to disorder or details of dielectric screening. 

The gap we measure is equivalent to the particle-hole excitation energy \cite{foutty_tunable_2023}. Mapping the field dependence of these gaps thus allows us to determine the spin character of charge (hole) excitations. To address how different excitations evolve in a field, we consider the effects of Coulomb interactions on the spin-split LLs in WSe$_2$, beyond the general effect of a density-dependent spin susceptibility. Using Hartree-Fock calculations with RPA-screened interactions which take into account the large LL mixing in this material, we study the charge gaps from the highest-energy filled spin-majority LL [identified by its orbital index $N$ and spin as ($|\nu|-1,\uparrow$)] to both the subsequent spin-majority ($|\nu|,\uparrow$) and lowest energy spin-minority ($1,\downarrow$) LLs (Supplementary Sec. 6-7). The resulting gaps are plotted in Fig.~\ref{fig:Fig2}f-j, with a schematic illustration of how the LL energies evolve with magnetic field in Fig. \ref{fig:Fig2}l.

The gap to the next unoccupied spin-majority LL is mostly set by the cyclotron energy, as exchange interactions affect both majority-spin LLs similarly. The gap to the lowest spin-minority LL (the `spin-flip gap') is determined by both single-particle and exchange interactions; the latter strongly renormalize this gap because they have different effects on the particle and hole excitations. The relative balance of these two contributions at a given filling factor will vary as the magnetic field (and therefore carrier density) is tuned. At low magnetic fields, exchange interactions are comparatively stronger and disfavor minority spin occupation, increasing the spin-flip gap. At higher magnetic fields, the kinetic energy dominates the behavior, leading to a linear decrease with $B$ from the large orbital mismatch between the relevant LLs. The result is a non-monotonic dependence of the spin-flip gap with $B$ so that it becomes competitive (and is eventually favored) compared with the cyclotron gap (Supplementary Sec. 7).

Our numerical calculations (Fig.~\ref{fig:Fig2}f-j) indicate that for the LLs we probe in our experiment, the curvature of the spin-flip gap is quite low at the crossover field $B_T$. This qualitatively matches the plateaus we observe over an intermediate field range in our measurements. We therefore interpret the transition region as a spin-polarized ground state that favors occupation of an opposite spin LL upon doping. At higher magnetic fields ($B > B_C$), the system undergoes a first-order transition to a mixed regime where the $(1, \downarrow)$ LL jumps to lower energy than the $(|\nu| -1, \uparrow)$ state. The precise ordering of LLs (for example, whether the $(1,\downarrow)$ LL also jumps below the $(|\nu|-2, \uparrow)$ state) is sensitive to details of the approximation in our theoretical calculations (Supplementary Sec. 7) and is ambiguous in experiment, so we restrict our quantitative comparison to $B < B_C$ in Fig. 2.

\section{Re-entrant magnetism and full spin phase diagram}

The close competition between distinct phases also affects the behavior of the system at partial LL filling. Our measurements near each LL phase transition reveal a sharp negative compressibility feature emanating outward towards lower hole density as the field decreases (Fig. \ref{fig:Fig3}a-c). We interpret this behavior, indicative of a first-order isospin phase transition \cite{eisenstein_negative_1992,feldman_fractional_2013,zhou_half-_2021,yu_correlated_2022}, as an extension of the LL reordering into compressible states of a partially filled LL (Fig. \ref{fig:Fig3}d). 

As holes are initially depleted from integer filling (pink region, Fig.~\ref{fig:Fig3}d), holes are removed from the highest energy majority spin LL (Fig. \ref{fig:Fig3}e). This depletion will decrease the exchange interactions and the partially-filed LL will be pushed towards the unoccupied minority spin LL. As additional holes are removed and the sample enters the gray region in Fig.~\ref{fig:Fig3}d, there is an abrupt reordering of spins and the minority spin LL is instead occupied (Fig. \ref{fig:Fig3}f). Similar phenomenology was also suggested by recent transport measurements in a related system \cite{shih_spin-selective_2023}. These transitions at partial LL filling, along with the observation of multiple LL gap plateaus at fixed $n$ (varying $B$), imply re-entrant spin polarization as the system sequentially fills, depletes, and again fills holes into the $N=1$ minority spin LL. This leads to a `sawtooth' boundary between fully and partially spin-polarized phases in the $n$-$B$ plane, which we show in the full spin phase diagram in Fig.~\ref{fig:Fig3}g. 

\begin{figure}[t!]
    \renewcommand{\thefigure}{\arabic{figure}}
    \centering
    \includegraphics[scale =1.0]{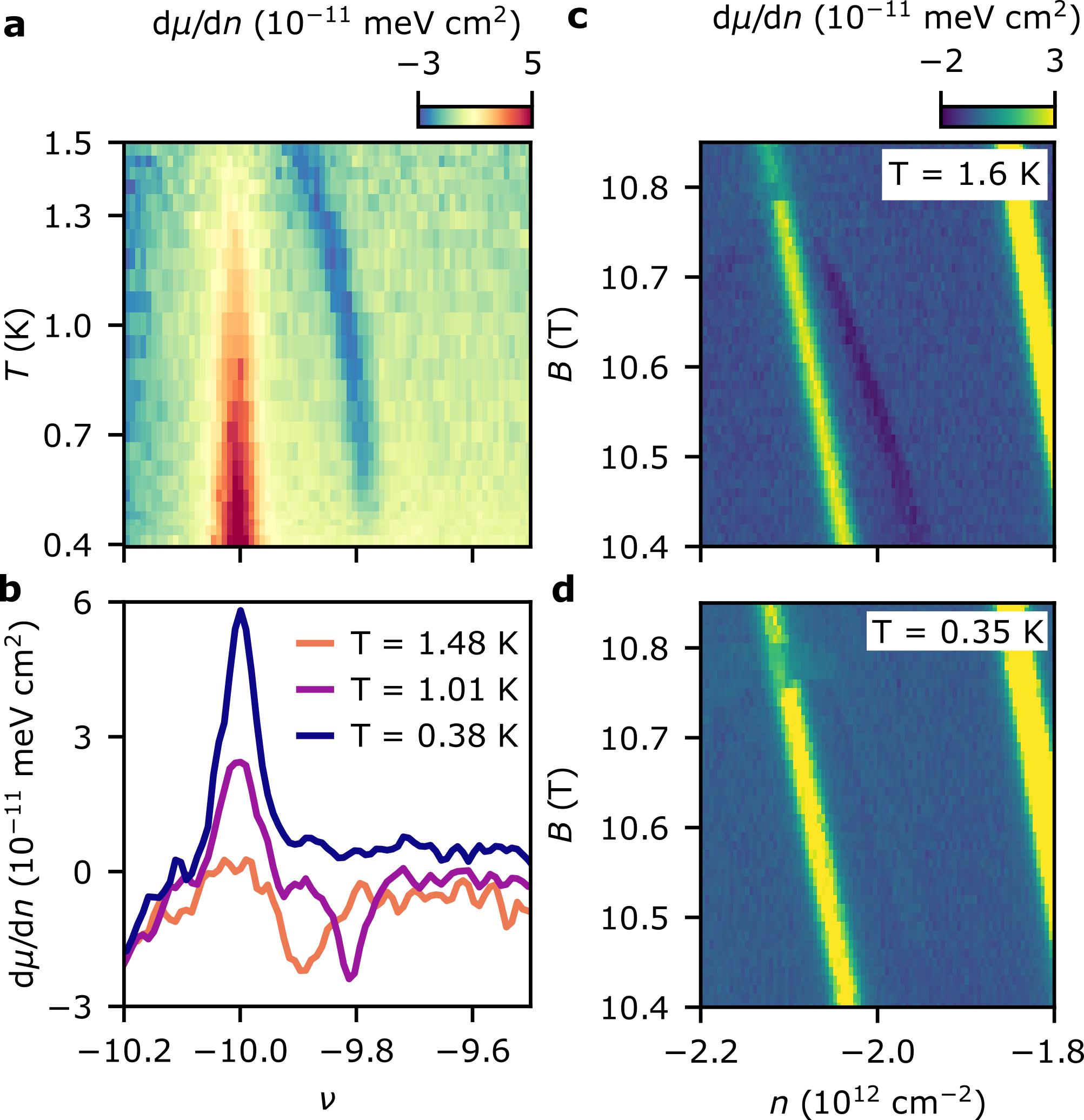}
        \caption{\textbf{Temperature dependence of the spin phase transition} \textbf{a}, d$\mu$/d$n$ as a function of $T$ and $\nu$ at $B = 4.85$ T. As the temperature decreases, the LL gap becomes stronger and the negative compressibility moves further away from integer filling and weakens at the lowest temperatures. \textbf{b}, Linecuts of d$\mu$/d$n$ from panel \textbf{a} at select temperatures. \textbf{c-d}, d$\mu$/d$n$ around the $\nu= -8$ transition at $T = 1.6$ K (\textbf{c}) and $T = 0.35$ K (\textbf{d}).}

    \label{fig:Fig4}
\end{figure}

Finally, we discuss how these phase transitions depend on temperature, which provides further insight into the relative free energies of distinct states. In Fig. \ref{fig:Fig4}a-b, we show d$\mu$/d$n$ as a function of $\nu$ and temperature $T$ at a constant magnetic field $B = 4.85$ T (near the $\nu = -10$ transition shown in Fig. \ref{fig:Fig3}b). The negative compressibility feature shifts to lower hole density as the system cools between $T = 1.5$ K and $T = 0.35$ K, indicating that the mixed phase is favored at higher temperatures and thus carries higher relative entropy. 

The incompressible LL gap significantly strengthens at lower temperatures, as expected. In contrast, the negative compressibility weakens at the lowest temperatures of our measurement, displaying a nonmonotonic magnitude as a function of temperature. This contrasts with measurements of isospin transitions in distinct systems, where such features sharpen at lower temperatures \cite{feldman_fractional_2013,zhou_half-_2021,yu_correlated_2022,kometter_hofstadter_2023,foutty_mapping_2023}. In Fig. \ref{fig:Fig4}c-d, we compare d$\mu$/d$n$ around the $\nu = -8$ transition at $T = 1.6$ K and $T = 0.35$ K, demonstrating that negative compressibility is barely visible at the lowest temperatures of our measurement. To explain this behavior, we use a Sommerfeld expansion to obtain a phenomenological model for the free energy around the phase transition. We find that at sufficiently low $T$, the relative slopes of the free energy as a function of density will be closer together due to an asymmetry in the density of states of the two phases, suppressing the negative d$\mu$/d$n$ at the phase transition (Supplementary Sec. 9).  

\section{Outlook}

In conclusion, our experiments reveal singular changes in the interaction-induced renormalization of LL energies at the crossover between spin-polarized and mixed states. The intertwined electronic and magnetic structure in this system enables gate control over macroscopic changes in magnetization at the Fermi level. Our results are relevant to a wider class of systems where many-body effects are even more prominent. While the sizeable single-particle Zeeman energies characteristic of monolayer WSe$_2$ makes the system susceptible to spin-polarization even without interactions, related systems have displayed exchange driven polarization in the absence of a large 
Zeeman energy \cite{roch_spin-polarized_2019,leisgang_exchange_2023}. Interaction-induced polarization (and related phase transitions) is also relevant within moir\'e heterostructures, in which flat moir\'e bands quench the kinetic energy akin to LLs and complete spin polarization can be favored at both zero and finite magnetic field \cite{zondiner_cascade_2020,saito_isospin_2021,yu_correlated_2022,anderson_programming_2023,kometter_hofstadter_2023}. Our comprehensive understanding of the relative kinetic and interaction effects at transitions between full and partial spin polarization provides a framework for both experimental and theoretical study of energetics within these still more strongly interacting platforms. 



\begin{thebibliography}{10}
\expandafter\ifx\csname url\endcsname\relax
  \def\url#1{\texttt{#1}}\fi
\expandafter\ifx\csname urlprefix\endcsname\relax\def\urlprefix{URL }\fi
\providecommand{\bibinfo}[2]{#2}
\providecommand{\eprint}[2][]{\url{#2}}

\bibitem{sondhi_skyrmions_1993}
\bibinfo{author}{Sondhi, S.~L.}, \bibinfo{author}{Karlhede, A.},
  \bibinfo{author}{Kivelson, S.~A.} \& \bibinfo{author}{Rezayi, E.~H.}
\newblock \bibinfo{title}{Skyrmions and the crossover from the integer to
  fractional quantum {Hall} effect at small {Zeeman} energies}.
\newblock \emph{\bibinfo{journal}{Physical Review B}}
  \textbf{\bibinfo{volume}{47}}, \bibinfo{pages}{16419--16426}
  (\bibinfo{year}{1993}).
\newblock \urlprefix\url{https://link.aps.org/doi/10.1103/PhysRevB.47.16419}.
\newblock \bibinfo{note}{Publisher: American Physical Society}.

\bibitem{piazza_first-order_1999}
\bibinfo{author}{Piazza, V.} \emph{et~al.}
\newblock \bibinfo{title}{First-order phase transitions in a quantum {Hall}
  ferromagnet}.
\newblock \emph{\bibinfo{journal}{Nature}} \textbf{\bibinfo{volume}{402}},
  \bibinfo{pages}{638--641} (\bibinfo{year}{1999}).
\newblock \urlprefix\url{https://www.nature.com/articles/45189}.
\newblock \bibinfo{note}{Number: 6762 Publisher: Nature Publishing Group}.

\bibitem{tutuc_-plane_2001}
\bibinfo{author}{Tutuc, E.}, \bibinfo{author}{De~Poortere, E.~P.},
  \bibinfo{author}{Papadakis, S.~J.} \& \bibinfo{author}{Shayegan, M.}
\newblock \bibinfo{title}{In-{Plane} {Magnetic} {Field}-{Induced} {Spin}
  {Polarization} and {Transition} to {Insulating} {Behavior} in
  {Two}-{Dimensional} {Hole} {Systems}}.
\newblock \emph{\bibinfo{journal}{Physical Review Letters}}
  \textbf{\bibinfo{volume}{86}}, \bibinfo{pages}{2858--2861}
  (\bibinfo{year}{2001}).
\newblock \urlprefix\url{https://link.aps.org/doi/10.1103/PhysRevLett.86.2858}.

\bibitem{gunawan_valley_2006}
\bibinfo{author}{Gunawan, O.} \emph{et~al.}
\newblock \bibinfo{title}{Valley {Susceptibility} of an {Interacting}
  {Two}-{Dimensional} {Electron} {System}}.
\newblock \emph{\bibinfo{journal}{Physical Review Letters}}
  \textbf{\bibinfo{volume}{97}}, \bibinfo{pages}{186404}
  (\bibinfo{year}{2006}).
\newblock
  \urlprefix\url{https://link.aps.org/doi/10.1103/PhysRevLett.97.186404}.

\bibitem{nomura_quantum_2006}
\bibinfo{author}{Nomura, K.} \& \bibinfo{author}{MacDonald, A.~H.}
\newblock \bibinfo{title}{Quantum {Hall} {Ferromagnetism} in {Graphene}}.
\newblock \emph{\bibinfo{journal}{Physical Review Letters}}
  \textbf{\bibinfo{volume}{96}}, \bibinfo{pages}{256602}
  (\bibinfo{year}{2006}).
\newblock
  \urlprefix\url{https://link.aps.org/doi/10.1103/PhysRevLett.96.256602}.
\newblock \bibinfo{note}{Publisher: American Physical Society}.

\bibitem{kott_valley-degenerate_2014}
\bibinfo{author}{Kott, T.~M.}, \bibinfo{author}{Hu, B.},
  \bibinfo{author}{Brown, S.~H.} \& \bibinfo{author}{Kane, B.~E.}
\newblock \bibinfo{title}{Valley-degenerate two-dimensional electrons in the
  lowest {Landau} level}.
\newblock \emph{\bibinfo{journal}{Physical Review B}}
  \textbf{\bibinfo{volume}{89}}, \bibinfo{pages}{041107}
  (\bibinfo{year}{2014}).
\newblock \urlprefix\url{https://link.aps.org/doi/10.1103/PhysRevB.89.041107}.

\bibitem{hunt_direct_2017}
\bibinfo{author}{Hunt, B.~M.} \emph{et~al.}
\newblock \bibinfo{title}{Direct measurement of discrete valley and orbital
  quantum numbers in bilayer graphene}.
\newblock \emph{\bibinfo{journal}{Nature Communications}}
  \textbf{\bibinfo{volume}{8}}, \bibinfo{pages}{948} (\bibinfo{year}{2017}).
\newblock \urlprefix\url{https://www.nature.com/articles/s41467-017-00824-w}.
\newblock \bibinfo{note}{Number: 1 Publisher: Nature Publishing Group}.

\bibitem{shi_bilayer_2022}
\bibinfo{author}{Shi, Q.} \emph{et~al.}
\newblock \bibinfo{title}{Bilayer {WSe2} as a natural platform for interlayer
  exciton condensates in the strong coupling limit}.
\newblock \emph{\bibinfo{journal}{Nature Nanotechnology}}
  \textbf{\bibinfo{volume}{17}}, \bibinfo{pages}{577--582}
  (\bibinfo{year}{2022}).
\newblock \urlprefix\url{https://www.nature.com/articles/s41565-022-01104-5}.
\newblock \bibinfo{note}{Number: 6 Publisher: Nature Publishing Group}.

\bibitem{shi_odd-_2020}
\bibinfo{author}{Shi, Q.} \emph{et~al.}
\newblock \bibinfo{title}{Odd- and even-denominator fractional quantum {Hall}
  states in monolayer {WSe2}}.
\newblock \emph{\bibinfo{journal}{Nature Nanotechnology}}
  \textbf{\bibinfo{volume}{15}}, \bibinfo{pages}{569--573}
  (\bibinfo{year}{2020}).
\newblock \urlprefix\url{https://www.nature.com/articles/s41565-020-0685-6}.
\newblock \bibinfo{note}{Number: 7 Publisher: Nature Publishing Group}.

\bibitem{maryenko_polarization-dependent_2014}
\bibinfo{author}{Maryenko, D.}, \bibinfo{author}{Falson, J.},
  \bibinfo{author}{Kozuka, Y.}, \bibinfo{author}{Tsukazaki, A.} \&
  \bibinfo{author}{Kawasaki, M.}
\newblock \bibinfo{title}{Polarization-dependent {Landau} level crossing in a
  two-dimensional electron system in a {MgZnO}/{ZnO} heterostructure}.
\newblock \emph{\bibinfo{journal}{Physical Review B}}
  \textbf{\bibinfo{volume}{90}}, \bibinfo{pages}{245303}
  (\bibinfo{year}{2014}).
\newblock \urlprefix\url{https://link.aps.org/doi/10.1103/PhysRevB.90.245303}.

\bibitem{falson_review_2018}
\bibinfo{author}{Falson, J.} \& \bibinfo{author}{Kawasaki, M.}
\newblock \bibinfo{title}{A review of the quantum {Hall} effects in
  {MgZnO}/{ZnO} heterostructures}.
\newblock \emph{\bibinfo{journal}{Reports on Progress in Physics}}
  \textbf{\bibinfo{volume}{81}}, \bibinfo{pages}{056501}
  (\bibinfo{year}{2018}).
\newblock \urlprefix\url{https://dx.doi.org/10.1088/1361-6633/aaa978}.
\newblock \bibinfo{note}{Publisher: IOP Publishing}.

\bibitem{xu_spin_2014}
\bibinfo{author}{Xu, X.}, \bibinfo{author}{Yao, W.}, \bibinfo{author}{Xiao, D.}
  \& \bibinfo{author}{Heinz, T.~F.}
\newblock \bibinfo{title}{Spin and pseudospins in layered transition metal
  dichalcogenides}.
\newblock \emph{\bibinfo{journal}{Nature Physics}}
  \textbf{\bibinfo{volume}{10}}, \bibinfo{pages}{343--350}
  (\bibinfo{year}{2014}).
\newblock \urlprefix\url{https://www.nature.com/articles/nphys2942}.
\newblock \bibinfo{note}{Number: 5 Publisher: Nature Publishing Group}.

\bibitem{gustafsson_ambipolar_2018}
\bibinfo{author}{Gustafsson, M.~V.} \emph{et~al.}
\newblock \bibinfo{title}{Ambipolar {Landau} levels and strong band-selective
  carrier interactions in monolayer {WSe2}}.
\newblock \emph{\bibinfo{journal}{Nature Materials}}
  \textbf{\bibinfo{volume}{17}}, \bibinfo{pages}{411--415}
  (\bibinfo{year}{2018}).
\newblock \urlprefix\url{https://www.nature.com/articles/s41563-018-0036-2}.

\bibitem{wang_valley-_2017}
\bibinfo{author}{Wang, Z.}, \bibinfo{author}{Shan, J.} \& \bibinfo{author}{Mak,
  K.~F.}
\newblock \bibinfo{title}{Valley- and spin-polarized {Landau} levels in
  monolayer {WSe2}}.
\newblock \emph{\bibinfo{journal}{Nature Nanotechnology}}
  \textbf{\bibinfo{volume}{12}}, \bibinfo{pages}{144--149}
  (\bibinfo{year}{2017}).
\newblock \urlprefix\url{https://www.nature.com/articles/nnano.2016.213}.
\newblock \bibinfo{note}{Number: 2 Publisher: Nature Publishing Group}.

\bibitem{movva_density-dependent_2017}
\bibinfo{author}{Movva, H.~C.} \emph{et~al.}
\newblock \bibinfo{title}{Density-{Dependent} {Quantum} {Hall} {States} and
  {Zeeman} {Splitting} in {Monolayer} and {Bilayer}
  \$\{{\textbackslash}mathrm\{{WSe}\}\}\_\{2\}\$}.
\newblock \emph{\bibinfo{journal}{Physical Review Letters}}
  \textbf{\bibinfo{volume}{118}}, \bibinfo{pages}{247701}
  (\bibinfo{year}{2017}).
\newblock
  \urlprefix\url{https://link.aps.org/doi/10.1103/PhysRevLett.118.247701}.
\newblock \bibinfo{note}{Publisher: American Physical Society}.

\bibitem{larentis_large_2018}
\bibinfo{author}{Larentis, S.} \emph{et~al.}
\newblock \bibinfo{title}{Large effective mass and interaction-enhanced
  {Zeeman} splitting of {K} -valley electrons in {MoSe} 2}.
\newblock \emph{\bibinfo{journal}{Physical Review B}}
  \textbf{\bibinfo{volume}{97}}, \bibinfo{pages}{201407}
  (\bibinfo{year}{2018}).
\newblock \urlprefix\url{https://link.aps.org/doi/10.1103/PhysRevB.97.201407}.

\bibitem{pisoni_interactions_2018}
\bibinfo{author}{Pisoni, R.} \emph{et~al.}
\newblock \bibinfo{title}{Interactions and {Magnetotransport} through
  {Spin}-{Valley} {Coupled} {Landau} {Levels} in {Monolayer}
  \$\{{\textbackslash}mathrm\{{MoS}\}\}\_\{2\}\$}.
\newblock \emph{\bibinfo{journal}{Physical Review Letters}}
  \textbf{\bibinfo{volume}{121}}, \bibinfo{pages}{247701}
  (\bibinfo{year}{2018}).
\newblock
  \urlprefix\url{https://link.aps.org/doi/10.1103/PhysRevLett.121.247701}.
\newblock \bibinfo{note}{Publisher: American Physical Society}.

\bibitem{li_spontaneous_2020}
\bibinfo{author}{Li, J.} \emph{et~al.}
\newblock \bibinfo{title}{Spontaneous {Valley} {Polarization} of {Interacting}
  {Carriers} in a {Monolayer} {Semiconductor}}.
\newblock \emph{\bibinfo{journal}{Physical Review Letters}}
  \textbf{\bibinfo{volume}{125}}, \bibinfo{pages}{147602}
  (\bibinfo{year}{2020}).
\newblock
  \urlprefix\url{https://link.aps.org/doi/10.1103/PhysRevLett.125.147602}.
\newblock \bibinfo{note}{Publisher: American Physical Society}.

\bibitem{shih_spin-selective_2023}
\bibinfo{author}{Shih, E.-M.} \emph{et~al.}
\newblock \bibinfo{title}{Spin-selective magneto-conductivity in {WSe}\$\_2\$}
  (\bibinfo{year}{2023}).
\newblock \urlprefix\url{http://arxiv.org/abs/2307.00446}.
\newblock \bibinfo{note}{ArXiv:2307.00446 [cond-mat]}.

\bibitem{pack_charge-transfer_2023}
\bibinfo{author}{Pack, J.} \emph{et~al.}
\newblock \bibinfo{title}{Charge-transfer {Contact} to a {High}-{Mobility}
  {Monolayer} {Semiconductor}} (\bibinfo{year}{2023}).
\newblock \urlprefix\url{http://arxiv.org/abs/2310.19782}.
\newblock \bibinfo{note}{ArXiv:2310.19782 [cond-mat]}.

\bibitem{vakili_spin_2004}
\bibinfo{author}{Vakili, K.}, \bibinfo{author}{Shkolnikov, Y.~P.},
  \bibinfo{author}{Tutuc, E.}, \bibinfo{author}{De~Poortere, E.~P.} \&
  \bibinfo{author}{Shayegan, M.}
\newblock \bibinfo{title}{Spin {Susceptibility} of {Two}-{Dimensional}
  {Electrons} in {Narrow} {AlAs} {Quantum} {Wells}}.
\newblock \emph{\bibinfo{journal}{Physical Review Letters}}
  \textbf{\bibinfo{volume}{92}}, \bibinfo{pages}{226401}
  (\bibinfo{year}{2004}).
\newblock
  \urlprefix\url{https://link.aps.org/doi/10.1103/PhysRevLett.92.226401}.

\bibitem{attaccalite_correlation_2002}
\bibinfo{author}{Attaccalite, C.}, \bibinfo{author}{Moroni, S.},
  \bibinfo{author}{Gori-Giorgi, P.} \& \bibinfo{author}{Bachelet, G.~B.}
\newblock \bibinfo{title}{Correlation {Energy} and {Spin} {Polarization} in the
  {2D} {Electron} {Gas}}.
\newblock \emph{\bibinfo{journal}{Physical Review Letters}}
  \textbf{\bibinfo{volume}{88}}, \bibinfo{pages}{256601}
  (\bibinfo{year}{2002}).
\newblock
  \urlprefix\url{https://link.aps.org/doi/10.1103/PhysRevLett.88.256601}.

\bibitem{ando_theory_1974}
\bibinfo{author}{Ando, T.} \& \bibinfo{author}{Uemura, Y.}
\newblock \bibinfo{title}{Theory of {Quantum} {Transport} in a
  {Two}-{Dimensional} {Electron} {System} under {Magnetic} {Fields}. {I}.
  {Characteristics} of {Level} {Broadening} and {Transport} under {Strong}
  {Fields}}.
\newblock \emph{\bibinfo{journal}{Journal of the Physical Society of Japan}}
  \textbf{\bibinfo{volume}{36}}, \bibinfo{pages}{959--967}
  (\bibinfo{year}{1974}).
\newblock \urlprefix\url{https://journals.jps.jp/doi/10.1143/JPSJ.36.959}.

\bibitem{foutty_tunable_2023}
\bibinfo{author}{Foutty, B.~A.} \emph{et~al.}
\newblock \bibinfo{title}{Tunable spin and valley excitations of correlated
  insulators in $\Gamma$-valley moiré bands}.
\newblock \emph{\bibinfo{journal}{Nature Materials}} \bibinfo{pages}{1--6}
  (\bibinfo{year}{2023}).
\newblock \urlprefix\url{https://www.nature.com/articles/s41563-023-01534-z}.
\newblock \bibinfo{note}{Publisher: Nature Publishing Group}.

\bibitem{eisenstein_negative_1992}
\bibinfo{author}{Eisenstein, J.~P.}, \bibinfo{author}{Pfeiffer, L.~N.} \&
  \bibinfo{author}{West, K.~W.}
\newblock \bibinfo{title}{Negative compressibility of interacting
  two-dimensional electron and quasiparticle gases}.
\newblock \emph{\bibinfo{journal}{Physical Review Letters}}
  \textbf{\bibinfo{volume}{68}}, \bibinfo{pages}{674--677}
  (\bibinfo{year}{1992}).
\newblock \urlprefix\url{https://link.aps.org/doi/10.1103/PhysRevLett.68.674}.

\bibitem{feldman_fractional_2013}
\bibinfo{author}{Feldman, B.~E.} \emph{et~al.}
\newblock \bibinfo{title}{Fractional {Quantum} {Hall} {Phase} {Transitions} and
  {Four}-{Flux} {States} in {Graphene}}.
\newblock \emph{\bibinfo{journal}{Physical Review Letters}}
  \textbf{\bibinfo{volume}{111}}, \bibinfo{pages}{076802}
  (\bibinfo{year}{2013}).
\newblock
  \urlprefix\url{https://link.aps.org/doi/10.1103/PhysRevLett.111.076802}.
\newblock \bibinfo{note}{Publisher: American Physical Society}.

\bibitem{zhou_half-_2021}
\bibinfo{author}{Zhou, H.} \emph{et~al.}
\newblock \bibinfo{title}{Half- and quarter-metals in rhombohedral trilayer
  graphene}.
\newblock \emph{\bibinfo{journal}{Nature}} \textbf{\bibinfo{volume}{598}},
  \bibinfo{pages}{429--433} (\bibinfo{year}{2021}).
\newblock \urlprefix\url{https://www.nature.com/articles/s41586-021-03938-w}.
\newblock \bibinfo{note}{Number: 7881 Publisher: Nature Publishing Group}.

\bibitem{yu_correlated_2022}
\bibinfo{author}{Yu, J.} \emph{et~al.}
\newblock \bibinfo{title}{Correlated {Hofstadter} spectrum and flavour phase
  diagram in magic-angle twisted bilayer graphene}.
\newblock \emph{\bibinfo{journal}{Nature Physics}}
  \textbf{\bibinfo{volume}{18}}, \bibinfo{pages}{825--831}
  (\bibinfo{year}{2022}).
\newblock \urlprefix\url{https://www.nature.com/articles/s41567-022-01589-w}.
\newblock \bibinfo{note}{Number: 7 Publisher: Nature Publishing Group}.

\bibitem{kometter_hofstadter_2023}
\bibinfo{author}{Kometter, C.~R.} \emph{et~al.}
\newblock \bibinfo{title}{Hofstadter states and re-entrant charge order in a
  semiconductor moiré lattice}.
\newblock \emph{\bibinfo{journal}{Nature Physics}} \bibinfo{pages}{1--7}
  (\bibinfo{year}{2023}).
\newblock \urlprefix\url{https://www.nature.com/articles/s41567-023-02195-0}.
\newblock \bibinfo{note}{Publisher: Nature Publishing Group}.

\bibitem{foutty_mapping_2023}
\bibinfo{author}{Foutty, B.~A.} \emph{et~al.}
\newblock \bibinfo{title}{Mapping twist-tuned multi-band topology in bilayer
  {WSe}\$\_2\$} (\bibinfo{year}{2023}).
\newblock \urlprefix\url{http://arxiv.org/abs/2304.09808}.
\newblock \bibinfo{note}{ArXiv:2304.09808 [cond-mat]}.

\bibitem{roch_spin-polarized_2019}
\bibinfo{author}{Roch, J.~G.} \emph{et~al.}
\newblock \bibinfo{title}{Spin-polarized electrons in monolayer {MoS2}}.
\newblock \emph{\bibinfo{journal}{Nature Nanotechnology}}
  \textbf{\bibinfo{volume}{14}}, \bibinfo{pages}{432--436}
  (\bibinfo{year}{2019}).
\newblock \urlprefix\url{http://www.nature.com/articles/s41565-019-0397-y}.

\bibitem{leisgang_exchange_2023}
\bibinfo{author}{Leisgang, N.} \emph{et~al.}
\newblock \bibinfo{title}{Exchange energy of the ferromagnetic electronic
  ground-state in a monolayer semiconductor} (\bibinfo{year}{2023}).
\newblock \urlprefix\url{http://arxiv.org/abs/2311.02164}.
\newblock \bibinfo{note}{ArXiv:2311.02164 [cond-mat]}.

\bibitem{zondiner_cascade_2020}
\bibinfo{author}{Zondiner, U.} \emph{et~al.}
\newblock \bibinfo{title}{Cascade of phase transitions and {Dirac} revivals in
  magic-angle graphene}.
\newblock \emph{\bibinfo{journal}{Nature}} \textbf{\bibinfo{volume}{582}},
  \bibinfo{pages}{203--208} (\bibinfo{year}{2020}).
\newblock \urlprefix\url{https://www.nature.com/articles/s41586-020-2373-y}.

\bibitem{saito_isospin_2021}
\bibinfo{author}{Saito, Y.} \emph{et~al.}
\newblock \bibinfo{title}{Isospin {Pomeranchuk} effect in twisted bilayer
  graphene}.
\newblock \emph{\bibinfo{journal}{Nature}} \textbf{\bibinfo{volume}{592}},
  \bibinfo{pages}{220--224} (\bibinfo{year}{2021}).
\newblock \urlprefix\url{https://www.nature.com/articles/s41586-021-03409-2}.
\newblock \bibinfo{note}{Number: 7853 Publisher: Nature Publishing Group}.

\bibitem{anderson_programming_2023}
\bibinfo{author}{Anderson, E.} \emph{et~al.}
\newblock \bibinfo{title}{Programming correlated magnetic states with
  gate-controlled moiré geometry}.
\newblock \emph{\bibinfo{journal}{Science}} \textbf{\bibinfo{volume}{381}},
  \bibinfo{pages}{325--330} (\bibinfo{year}{2023}).
\newblock \urlprefix\url{https://www.science.org/doi/10.1126/science.adg4268}.
\newblock \bibinfo{note}{Publisher: American Association for the Advancement of
  Science}.

\bibitem{eisenstein_compressibility_1994}
\bibinfo{author}{Eisenstein, J.~P.}, \bibinfo{author}{Pfeiffer, L.~N.} \&
  \bibinfo{author}{West, K.~W.}
\newblock \bibinfo{title}{Compressibility of the two-dimensional electron gas:
  {Measurements} of the zero-field exchange energy and fractional quantum
  {Hall} gap}.
\newblock \emph{\bibinfo{journal}{Physical Review B}}
  \textbf{\bibinfo{volume}{50}}, \bibinfo{pages}{1760--1778}
  (\bibinfo{year}{1994}).
\newblock \urlprefix\url{https://link.aps.org/doi/10.1103/PhysRevB.50.1760}.

\end{thebibliography}


\section{Methods}
\subsection{Sample fabrication}
All WSe$_2$ devices were fabricated using standard dry transfer techniques, following a process identical to that of \cite{foutty_mapping_2023}. In brief, WSe$_2$ samples are fully encapsulated in hexagonal boron nitride (hBN) and gated via a graphite back gate. Contact to the WSe$_2$ is made via prepatterned Pt leads, which are locally gated by Cr/Au ``contact gates". All monolayers in this study were part of devices which also included twisted bilayers of WSe$_2$. Locations of measurements were chosen to be $> 500$ nm away from any ``stacking boundary" between monolayer and twisted bilayer portion of the samples.

\subsection{SET Measurements}
The SET sensor was fabricated by evaporating aluminum onto a pulled quartz rod, with an estimated diameter at the apex of $ 50 - 100$ nm. The SET ``tip" is brought to about $50$ nm above the sample surface. Scanning SET measurements were performed in a  Unisoku USM 1300 scanning probe microscope with a customized microscope head. a.c. excitations (2-5 mV peak-to-peak amplitude) were applied to both sample and back gate at distinct frequencies between 200 and 400 Hz. We then measure inverse compressibility $\textrm{d}\mu/\textrm{d}n \propto I_{\textrm{BG}} / I_{\textrm{2D}}$ where $ I_{\textrm{BG}}$ and $I_{\textrm{2D}}$ are measurements of the SET current demodulated at respective frequencies of the back gate and sample excitations \cite{yu_correlated_2022}. A d.c. offset voltage $V_{\textrm{2D}}$ is applied to the sample to maintain the working point of the SET at its maximum sensitivity point within a Coulomb blockade oscillation fringe chosen to be near the ``flat-band'' condition where the tip does not gate the sample, which minimizes tip-induced doping. The contact gates are held at a large, negative voltage throughout the measurement to maintain good electrical contact across variable hole doping. Measurements are taken at $T = 0.35$ K unless otherwise noted.

\subsection{Gap measurement}
The gap sizes shown in Fig. \ref{fig:Fig2} in the main text are measurements of the step in the chemical potential $\mu(n)$. Practically, this is extracted by numerically integrating the measured $\textrm{d}\mu/\textrm{d}n$ signal across the gap. To accurately measure the gap on top of slowly-varying negative compressibility coming from long-range interactions at low density, we subtract a small background before integrating, analogous to Refs. \cite{eisenstein_compressibility_1994,kometter_hofstadter_2023}. This background is taken from averaging the value of $\textrm{d}\mu/\textrm{d}n$ on either side of the gap, avoiding sharp negative compressibility features as in Fig. \ref{fig:Fig3}. This background subtraction also accounts for a.c. charging artifacts in the measurement of d$\mu$/d$n$ (e.g. the vertical features in Fig.~\ref{fig:Fig1}b), which can be independently identified because they are not present in direct measurements of $\mu(n)$ at d.c. timescales \cite{foutty_tunable_2023} and they depend on the voltage applied to the contact gates. In our measurements, the charging artifacts only have a constant additive effect to the background, which we confirm based on the consistency of the LL gaps as they cross through the artifacts. 

\section{Data availability}
The data that supports the findings of this study are available from the corresponding authors upon reasonable request.

\section{Code availability}
The codes that support the findings of this study are available from the corresponding authors upon reasonable request.

\section{Acknowledgements}
We acknowledge helpful conversations with Allan H. MacDonald and Brian Skinner. Device fabrication and scanning SET measurements were primarily supported by NSF-DMR-2103910. Supporting data from bilayer WSe$_2$ was supported by the Department of Energy, Office of Basic Energy Sciences, award number DE-SC0023109. V.C, Z.H., and S.A.K. were supported by Department of Energy, Office of Basic Energy Sciences, under Contract No. DE-AC02-76SF00515. K.W. and T.T. acknowledge support from the JSPS KAKENHI (Grant Numbers 20H00354 and 23H02052) and World Premier International Research Center Initiative (WPI), MEXT, Japan. B.A.F. acknowledges a Stanford Graduate Fellowship. Part of this work was performed at the Stanford Nano Shared Facilities (SNSF), supported by the National Science Foundation under award ECCS-2026822.

\section{Author contribution}
B.A.F and C.R.K. conducted the SET experiments. V.C. and Z.H. conducted theoretical calculations. B.A.F. fabricated the samples. S.L. and J.C.H. provided WSe$_2$ crystals. K.W. and T.T. provided hBN crystals. S.A.K and B.E.F. supervised the work. All authors participated in analysis of the data and writing of the manuscript.

\section{Competing interests}
The authors declare no competing interest. 

\end{document}


\title{Supplementary Material for:\\Anomalous Landau level gaps near magnetic transitions in monolayer WSe$_2$}

\author{Benjamin A. Foutty}
\affiliation{Geballe Laboratory for Advanced Materials, Stanford, CA 94305, USA}
\affiliation{Department of Physics, Stanford University, Stanford, CA 94305, USA}

\author{Vladimir Calvera}
\affiliation{Department of Physics, Stanford University, Stanford, CA 94305, USA}

\author{Zhaoyu Han}
\affiliation{Department of Physics, Stanford University, Stanford, CA 94305, USA}

\author{Carlos R. Kometter}
\affiliation{Geballe Laboratory for Advanced Materials, Stanford, CA 94305, USA}
\affiliation{Department of Physics, Stanford University, Stanford, CA 94305, USA}

\author{Song Liu}
\affiliation{Department of Mechanical Engineering, Columbia University, New York, NY, 10027, USA}

\author{Kenji Watanabe}
\affiliation{Research Center for Electronic and Optical Materials, National Institute for Materials Science, 1-1 Namiki, Tsukuba 305-0044, Japan}

\author{Takashi Taniguchi}
\affiliation{Research Center for Materials Nanoarchitectonics, National Institute for Materials Science, 1-1 Namiki, Tsukuba 305-0044, Japan}

\author{James C. Hone}
\affiliation{Department of Mechanical Engineering, Columbia University, New York, NY, 10027, USA}

\author{Steven A. Kivelson}
\affiliation{Department of Physics, Stanford University, Stanford, CA 94305, USA}

\author{Benjamin E. Feldman}
\email{bef@stanford.edu}
\affiliation{Geballe Laboratory for Advanced Materials, Stanford, CA 94305, USA}
\affiliation{Department of Physics, Stanford University, Stanford, CA 94305, USA}
\affiliation{Stanford Institute for Materials and Energy Sciences, SLAC National Accelerator Laboratory, Menlo Park, CA 94025, USA}

\maketitle
\tableofcontents

\section{Supporting data from additional samples}\label{sec:AdditionalSamples}

As mentioned in the main text, we measure three distinct monolayer WSe$_2$ samples, which we label Sample A, B, and C. All data in the main text is from a single location in Sample A, except for Fig. 4c-d, which is from Sample B. The starting WSe$_2$ material is from different sources and the samples have different distances to their respective back gates (set by thickness of the bottom hBN flake), which may affect dielectric screening. These are summarized in Table \ref{tab:Tab1}. As commented in the main text, we extract an effective mass $m^*$ from the linear field dependence of the LL gaps in each sample using the relation $\Delta_{\rm{cyc}} = \frac{\hbar e B}{m^*}$ where $\hbar$ is the reduced Planck constant and $e$ is the charge of the electron. The effective mass shows small variations across the three samples we studied, ranging from $m^* = 0.31m_e$ to 0.42$m_e$. This is slightly lower than prior measurements of LL gap sizes found but falls within the range of ARPES measurements and first-principles calculations of the WSe$_2$ valence band mass at $B = 0$. \cite{gustafsson_ambipolar_2018,wang_valley-_2017,le_spinorbit_2015,wilson_determination_nodate,liu_three-band_2013,kormanyos_kp_2015}. Differences in the amount of disorder in each sample leads to quantitative differences in the measured Landau level (LL) gaps: increasing disorder reduces gap size due to LL broadening. Additionally, sample-specific differences in dielectric breakdown upon gating and contact transparency produce different measurable density windows. For example, in Sample B, we are unable to measure at hole densities below $|n| = 1\times 10^{12}\ \rm{cm}^{-2}$, as the sample was unable to charge (on a.c. timescales) at those densities. Nonetheless, all of the qualitative features discussed in the main text are reproduced in multiple samples. Below, we detail further supporting data from additional devices.
\begin{table*}[h]
\renewcommand{\thetable}{S\arabic{table}}

\begin{center}
\begin{tabular}{ |c|c|c|c|} 
 \hline
 Sample & Source & Bottom hBN thickness (nm) & Measured effective mass via Landau level slope (in units of $\mZero$) \\  \hline
 Sample A & HQ Graphene & 44 & 0.31 $\pm$ 0.02 \\ \hline
 Sample B & HQ Graphene & 5.5 & 0.40 $\pm$ 0.01 \\  \hline 
 Sample C & Columbia & 11 & 0.42 $\pm$ 0.01  \\
 \hline
\end{tabular}
\end{center}
\caption{\textbf{Summary of different samples.}}
\label{tab:Tab1}
\end{table*}

\begin{figure*}[h]
    \renewcommand{\thefigure}{S\arabic{figure}}
    \centering
    \includegraphics[scale=1.0]{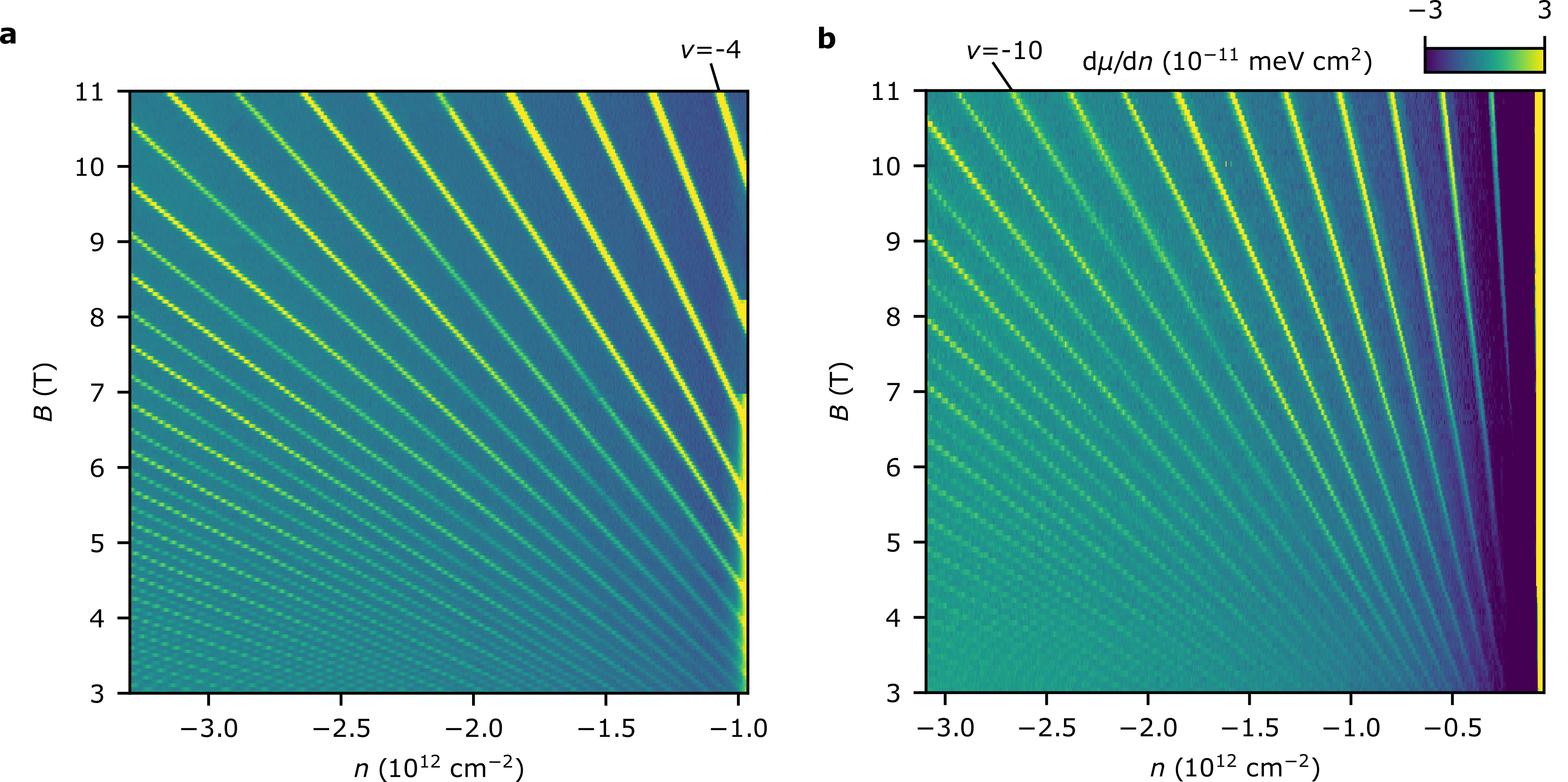}
    \caption{\textbf{Landau fans from distinct devices.} \textbf{a-b}, Landau fans of inverse electronic compressibility d$\mu$/d$n$ as a function of hole density $n$ and magnetic field $B$ from Sample B (\textbf{a}) and Sample C (\textbf{b}). Data in \textbf{a} is restricted to $|n| > 1\times 10^{12}$ cm$^{-2}$ due to insufficient charging at the location probed by the SET tip for this device.} 
    \label{fig:LFans}
\end{figure*}

In Fig. \ref{fig:LFans}, we show Landau fans for Samples B and C, which can be compared with the data in Fig. 1a. The densities of the phase transitions plotted in Fig. 1f are identified from these figures. The phase transitions in Sample B are consistently at (slightly) lower density than the other two samples, which is consistent with enhanced screening in the device (due to thinner hBN dielectric). For this reason, we observe the $\nu = -8$ transition in this device at $B = 10.8$ T (Fig. 4c-d) but not in Samples A and C up to the highest magnetic field ($B = 11$ T) available in our measurement system.

\begin{figure*}[h]
    \renewcommand{\thefigure}{S\arabic{figure}}
    \centering
    \includegraphics[scale=1.0]{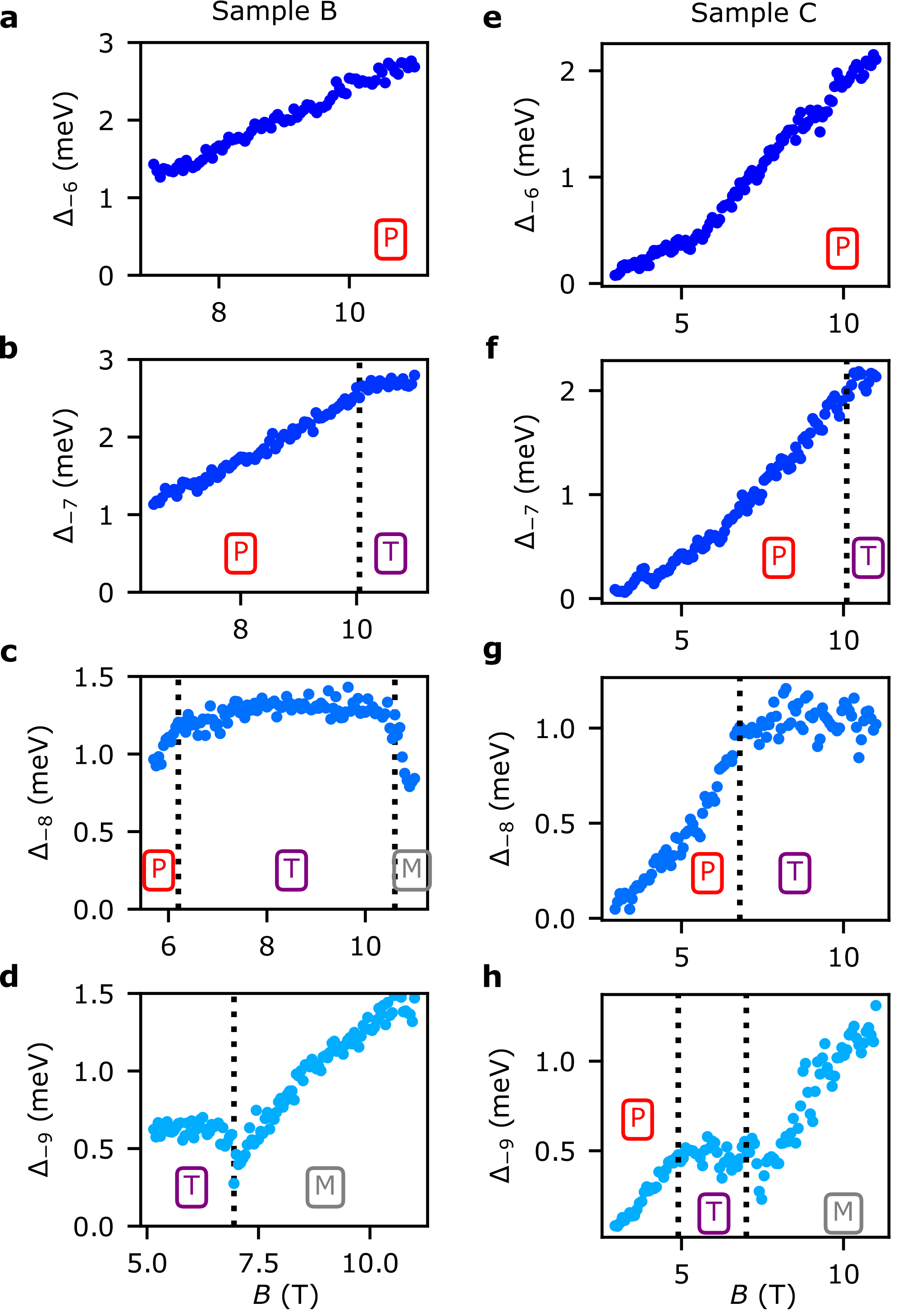}
    \caption{\textbf{Landau level (LL) gaps from distinct devices.} \textbf{a-h}, LL gaps $\Delta_{\nu}$ as a function of $B$ for a selection of filling factors $\nu$ from Sample B (\textbf{a-d}) and Sample C (\textbf{e-h}).} 
    \label{fig:SuppLLGaps}
\end{figure*}

In Fig. \ref{fig:SuppLLGaps}, we present a selection of LL gaps from Samples B and C, which qualitatively match the data presented in Fig. 2a-e. While the different levels of disorder in each sample changes the quantitative gap magnitudes at a given magnetic field, all devices exhibit plateaus in the LL gaps at $\nu = -7,-8,$ and $-9$. The gaps in Sample B are inaccessible at lower magnetic fields (precluding clear observation of the polarized regime for $\nu=-8$ and $-9$) because that sample can be measured only at higher densities.

\begin{figure*}[h]
    \renewcommand{\thefigure}{S\arabic{figure}}
    \centering
    \includegraphics[scale=1.0]{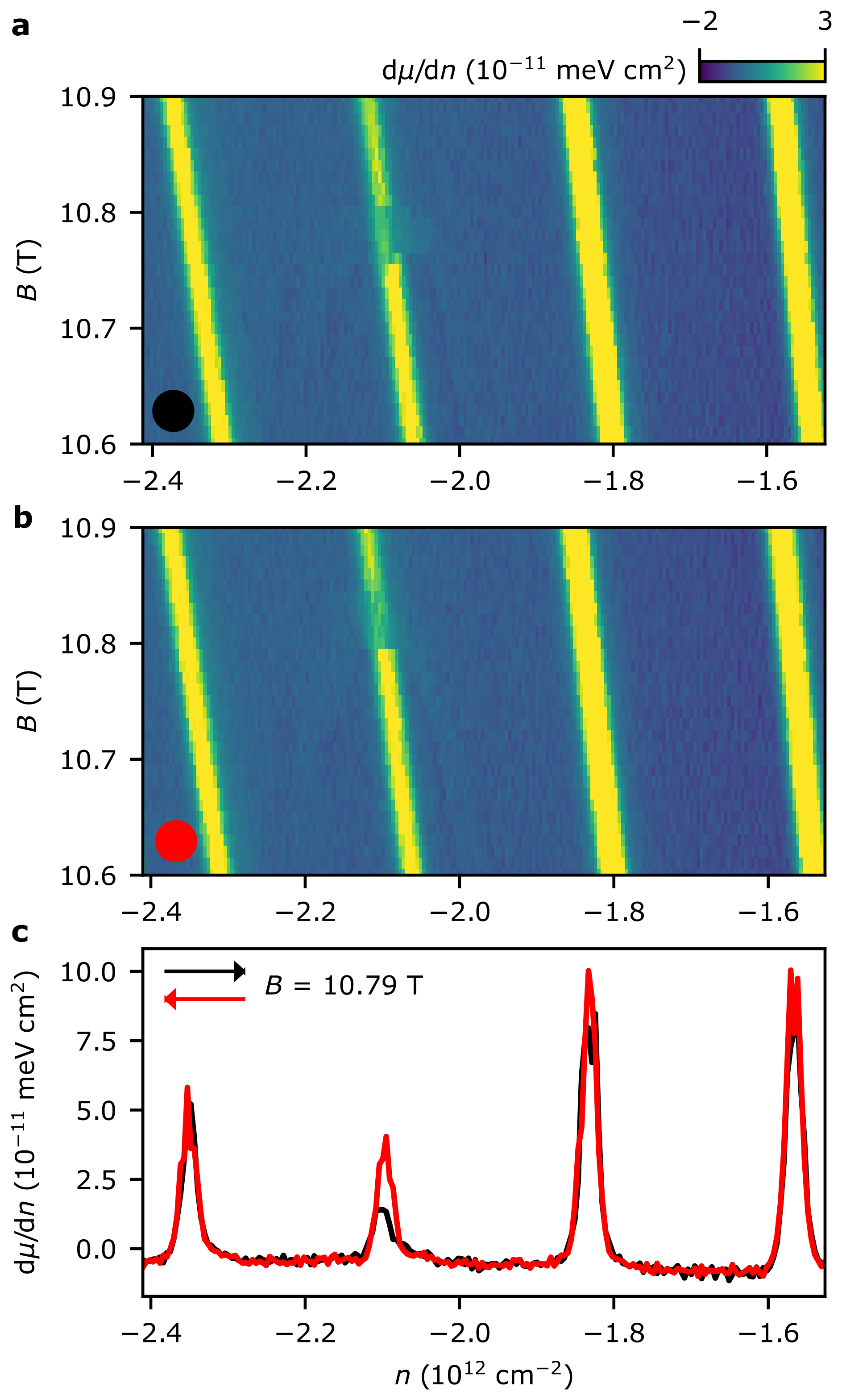}
    \caption{\textbf{Gate-tuned hysteresis in Sample B.} \textbf{a-b}, d$\mu$/d$n$ as a function of $n$ and $B$ around the $\nu = -8$ LL transition in Sample B. In \textbf{a} (\textbf{b}), the density is swept from left to right (right to left). Depending on the sweep direction, there is a difference of $B \approx 40$ mT in the magnetic field at which the LL gap drops sharply. \textbf{c}, Linecuts of d$\mu$/d$n$ as a function of $n$ at a constant magnetic field $B = 10.79$ T from panels $\textbf{a-b}$. The only notable difference is the magnitude of d$\mu$/d$n$ at $\nu = -8$ ($n\approx 2.1\times10^{12}$ cm$^{-2}$).} 
    \label{fig:SuppSampBHyst}
\end{figure*}

\begin{figure*}[h]
    \renewcommand{\thefigure}{S\arabic{figure}}
    \centering
    \includegraphics[scale=1.0]{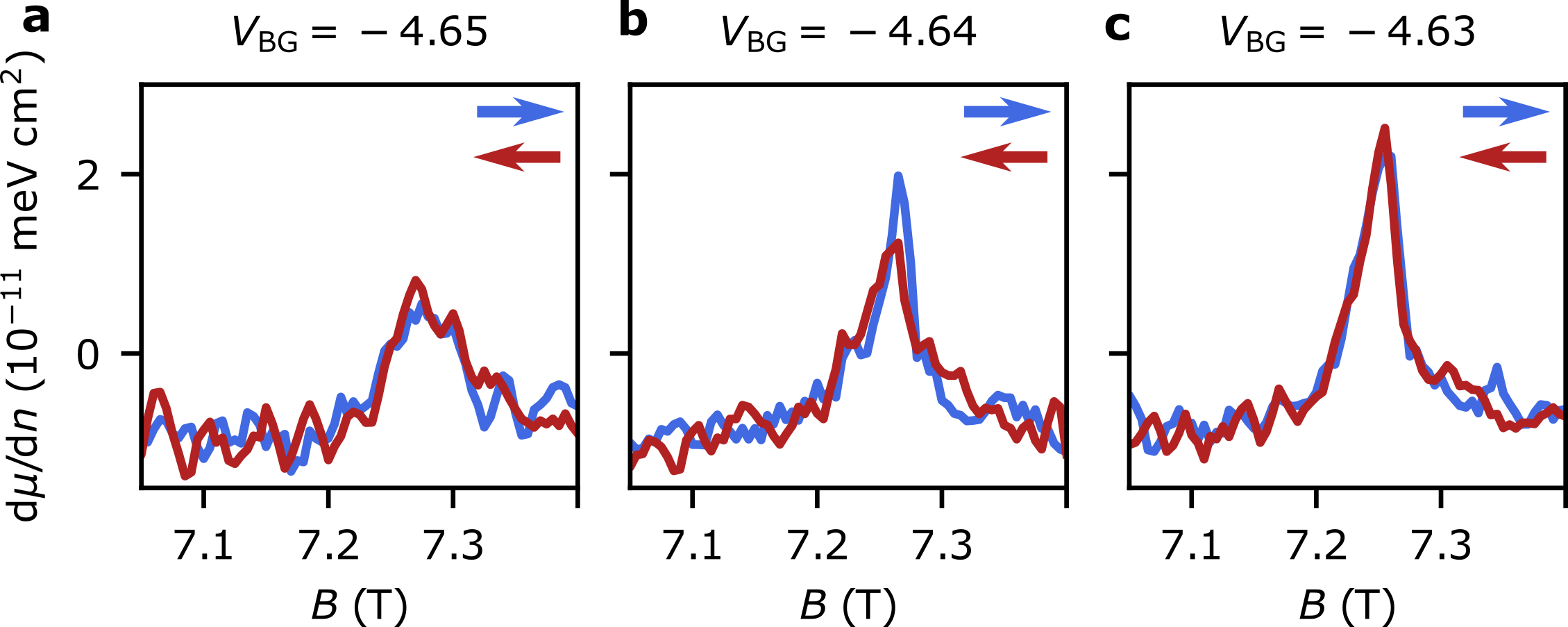}
    \caption{\textbf{Field-tuned hysteresis in Sample A.} \textbf{a-c}, d$\mu$/d$n$ as a function of $B$ (fast-axis) at three representative hole densities set by the back gate voltage $V_{\textrm{BG}}$ around the $\nu = -9$ phase transition. At the LL phase transition, we observe hysteresis while sweeping the magnetic field as the fast axis (\textbf{b}), while hysteresis is absent at nearby hole densities (\textbf{a},\textbf{c}). This data is taken at temperature $T = $ 1.6 K.} 
    \label{fig:BFieldHysteresis}
\end{figure*}

In Fig. \ref{fig:SuppSampBHyst} we present a secondary example of hysteresis at the transition into the mixed regime in Sample B. At $T = 0.35$ K, we observe a hysteresis of $B \approx 40$ mT at the $\nu = -8$ transition depending on which direction the density is swept. We also observe hysteresis in the magnetic field around these phase transitions (Fig. \ref{fig:BFieldHysteresis}), though we focus on density tuned hysteresis in the main text as our scanning probe set-up is more amenable to sweeping density as the fast axis. 

\section{Fractional quantum Hall states}\label{sec:FQH}
\begin{figure*}[h]
    \renewcommand{\thefigure}{S\arabic{figure}}
    \centering
    \includegraphics[scale=1.0]{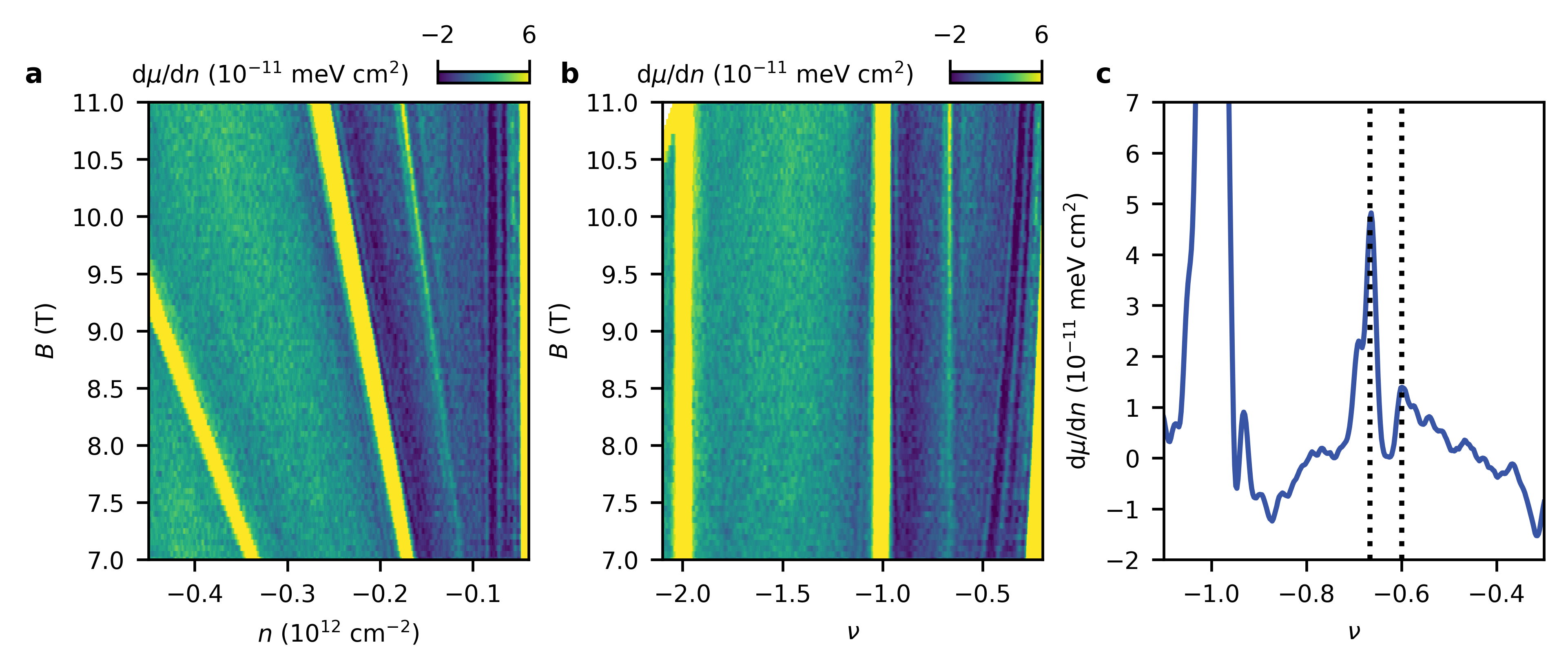}
    \caption{\textbf{Fractional quantum Hall states at high magnetic fields.} \textbf{a}, Landau fan of d$\mu$/d$n$ as a function of $n$ and $B$ in Sample A, showing a clear fractional quantum Hall state with slope $-2/3$. \textbf{b}, Data from panel \textbf{a} plotted as a function of filling factor $\nu$ and $B$. \textbf{c}, d$\mu$/d$n$ as a function of $\nu$ averaged between $B = 9.5$ T and $B = 11$ T. Dotted lines highlight an incompressible state at $\nu = -2/3$ and a weaker peak at $\nu = -3/5$.} 
    \label{fig:SuppFQH}
\end{figure*}

In Fig. \ref{fig:SuppFQH}, we present higher resolution data from Sample A at low densities and high fields, highlighting the fractional quantum Hall states in the lowest LL. We observe an incompressible state at $\nu = -\frac{2}{3}$ that persists down to $B = 7$ T, and a developing FQH state at $\nu = -\frac{3}{5}$ appearing around $B = 10$ T. The strongly particle-hole antisymmetric FQH in the lowest LL is likely due to the effect of strong LL mixing at these fields \cite{shi_odd-_2020}.

\section{Persistent LL gaps at crossings in the mixed regime}\label{sec:PersistentLLs}
\begin{figure*}[h]
    \renewcommand{\thefigure}{S\arabic{figure}}
    \centering
    \includegraphics[scale=1.0]{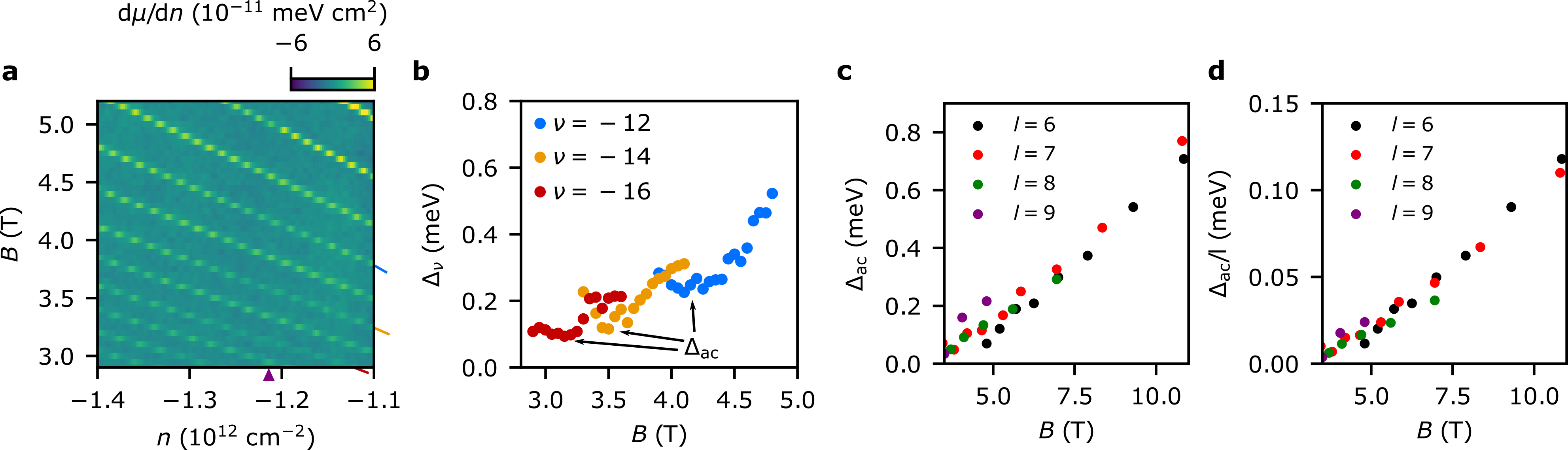}
    \caption{\textbf{Persistent gaps at LL crossings in the mixed regime.} \textbf{a}, d$\mu$/d$n$ as a function of $n$ and $B$ in the mixed regime. A crossover (marked by the purple triangle) occurs near $n = -1.2 \times 10^{12}$ cm$^{-2}$, where the even integer LL gaps exhibit a local minimum while odd integer gaps have a local maximum.  \textbf{b}, LL gaps $\Delta_{\nu}$ extracted from \textbf{a} for a selection of even integer LL gaps. All show local minima, which we denote as $\Delta_{\textrm{ac}}$, the `anti-crossing' gap size. \textbf{c}, The anti-crossing gaps $\Delta_{\textrm{ac}}$ plotted as a function of $B$, where $l$ describes the mismatch between the filling factors of each spin at the crossing point, as in \cite{de_poortere_resistance_2000}. \textbf{d}, Identical data to \textbf{c} with the $\Delta_{\textrm{ac}}$ rescaled by $1/l$.} 
    \label{fig:SuppACScaling}
\end{figure*}

In Fig. \ref{fig:SuppACScaling} we present further data on the gap sizes in the mixed LL regime, where there are a plethora of LL crossings due to the density dependence of the Zeeman energy \cite{gustafsson_ambipolar_2018,shi_odd-_2020}. As mentioned in the main text, there are no sharp changes in the behavior of these gaps (i.e. no obvious first-order transitions), and individual LL gaps smoothly decrease and increase. At sufficiently low temperatures, there are persistent gaps \cite{pisoni_interactions_2018} that remain open at these crossings (i.e., the thermodynamic gap does not go all the way to zero), as shown for a particular range of parameters in Fig. \ref{fig:SuppACScaling}a-b.  We present the minimum gaps for each measured LL ``anti-crossing" in Sample B in Fig. \ref{fig:SuppACScaling}c. Here, the index $l$ specifies the spin imbalance of occupied LLs at the given density of the crossing. For example, $l = 8$ crossings denote the crossings of LLs $(N_\uparrow, N_\downarrow) = (8,1)$ (closure of $\nu = -9$ gap), $(9,2)$, $(10,3)$, and so on. Practically, this is equivalent to grouping the crossings that occur at a given hole density. In prior measurements of quantum Hall ferromagnetism in AlAs, the gaps at the crossing points were noticed to scale as $1/l$ \cite{de_poortere_resistance_2000,vakili_dependence_2006}. In Fig. \ref{fig:SuppACScaling}d, we present the data from panel a scaled as $1/l$. The data are consistent with such a scaling. However, a conclusive scaling is difficult to determine because $l$ is quite high at experimentally accessible transitions (in comparison to AlAs). Therefore, this scaling will be less obvious, even if there is little apparent dependence on hole density \cite{pisoni_interactions_2018}. In an ideal case, a quantum Hall ferromagnetic gap should be set by the exchange energy, which goes as $\sqrt{B}$, rather than a linear dependence on $B$. However, LL mixing as well as disorder can reduce the quantum Hall ferromagnetic gap sizes and show a linear dependence on $B$ \cite{de_poortere_resistance_2000,ma_robust_2022}.

\section{Additional discussion of LL gaps}\label{sec:LLGapsvB}
\begin{figure*}[h]
    \renewcommand{\thefigure}{S\arabic{figure}}
    \centering
    \includegraphics[scale=1.0]{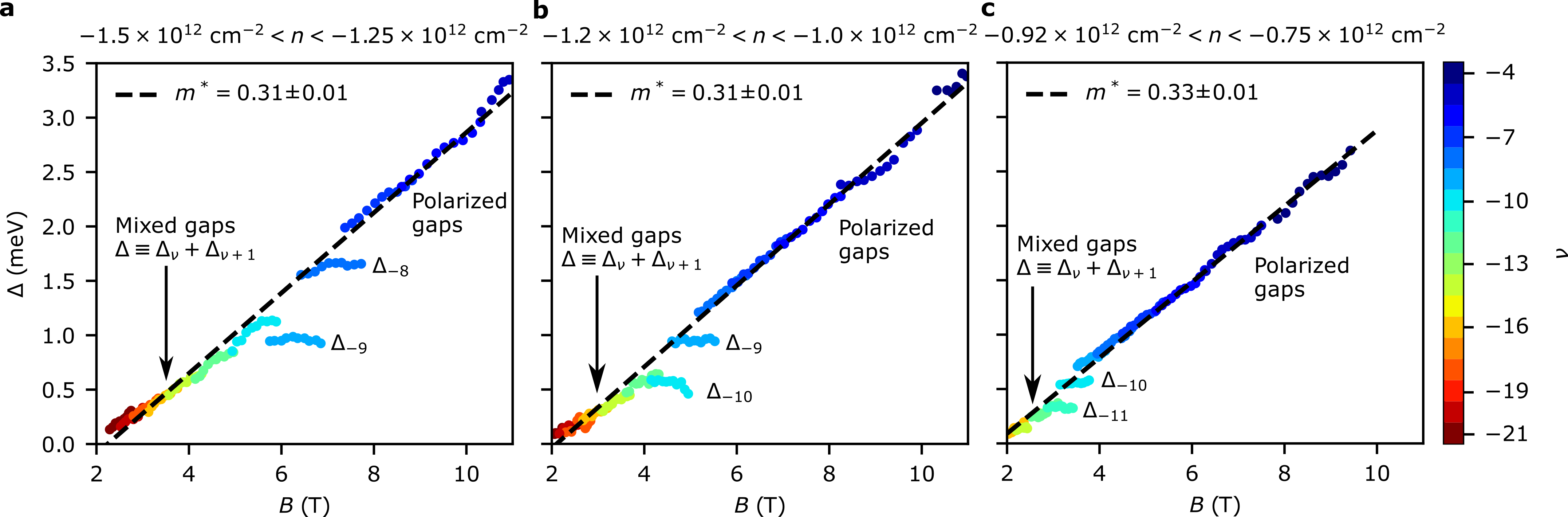}
    \caption{\textbf{LL gaps as a function of $B$ binned within different ranges of $n$.} \textbf{a}, Measured gaps $\Delta$ as a function of $B$ in Sample A in the range of $-1.5\times 10^{12}$ cm$^{-2} < n < -1.25\times 10^{12}$ cm$^{-2}$. Polarized and spin-flip gaps (from $\nu = -5$ to $\nu = -9$) correspond to $\Delta \equiv \Delta_{\nu}$, while mixed gaps $\nu \le -10$, are plotted as $\Delta \equiv \Delta_{\nu} + \Delta_{\nu +1}$ for odd $\nu$. The color of each point indicates the LL gap that is being plotted. \textbf{b}, Similar to \textbf{a}, but for the range of  $-1.2\times 10^{12}$ cm$^{-2} < n < -1.0\times 10^{12}$ cm$^{-2}$. Polarized and spin-flip gaps (from $\nu = -4$ to $\nu = -10$) correspond to $\Delta \equiv \Delta_{\nu}$, while mixed gaps $\nu \le -11$, are plotted as $\Delta \equiv \Delta_{\nu} + \Delta_{\nu +1}$ for even $\nu$. \textbf{c}, Similar to \textbf{a}, but for the range of  $-0.92\times 10^{12}$ cm$^{-2} < n < -0.75\times 10^{12}$ cm$^{-2}$. Polarized and spin-flip gaps (from $\nu = -4$ to $\nu = -11$) correspond to $\Delta \equiv \Delta_{\nu}$, while mixed gaps $\nu \le -12$, are plotted as $\Delta \equiv \Delta_{\nu} + \Delta_{\nu +1}$ for odd $\nu$. } 
    \label{fig:LLGapsvB}
\end{figure*}

In Fig. \ref{fig:LLGapsvB} we plot the experimentally measured LL gaps from Sample A in a representation similar to Fig. 2k in the main text at a number of distinct density `windows'. Here, data in each panel corresponds to a particular range of hole density, chosen to be between the densities of LL crossings in the mixed regime. For each polarized LL gap (including the ``transition" regime where spin-flip excitations are preferred), we plot its value against magnetic field measured within this range of hole density, where the color indicates the filling factor. For gaps in the mixed regime within this density range, we plot $\Delta_\nu + \Delta_{\nu - 1}$, starting from the lowest filling factor in the mixed regime. For example if $\nu = -10$ is the first mixed LL gap, we plot $\Delta_{-10} + \Delta_{-11}$, $\Delta_{-12} + \Delta_{-13}$, and so on. Plotted in this manner, it is clear that the polarized and mixed gaps (when the latter are summed appropriately) have roughly the same effective mass, matching the picture described in Refs.~\cite{gustafsson_ambipolar_2018,shi_odd-_2020}. Additionally, within each density window, there are exactly two LL gaps that plateau and lie below the expected cyclotron gap. 

\section{LLs in a naturally-stacked (Bernal) bilayer}\label{sec:BL_LLs}

\begin{figure*}[h]
    \renewcommand{\thefigure}{S\arabic{figure}}
    \centering
    \includegraphics[scale=1.0]{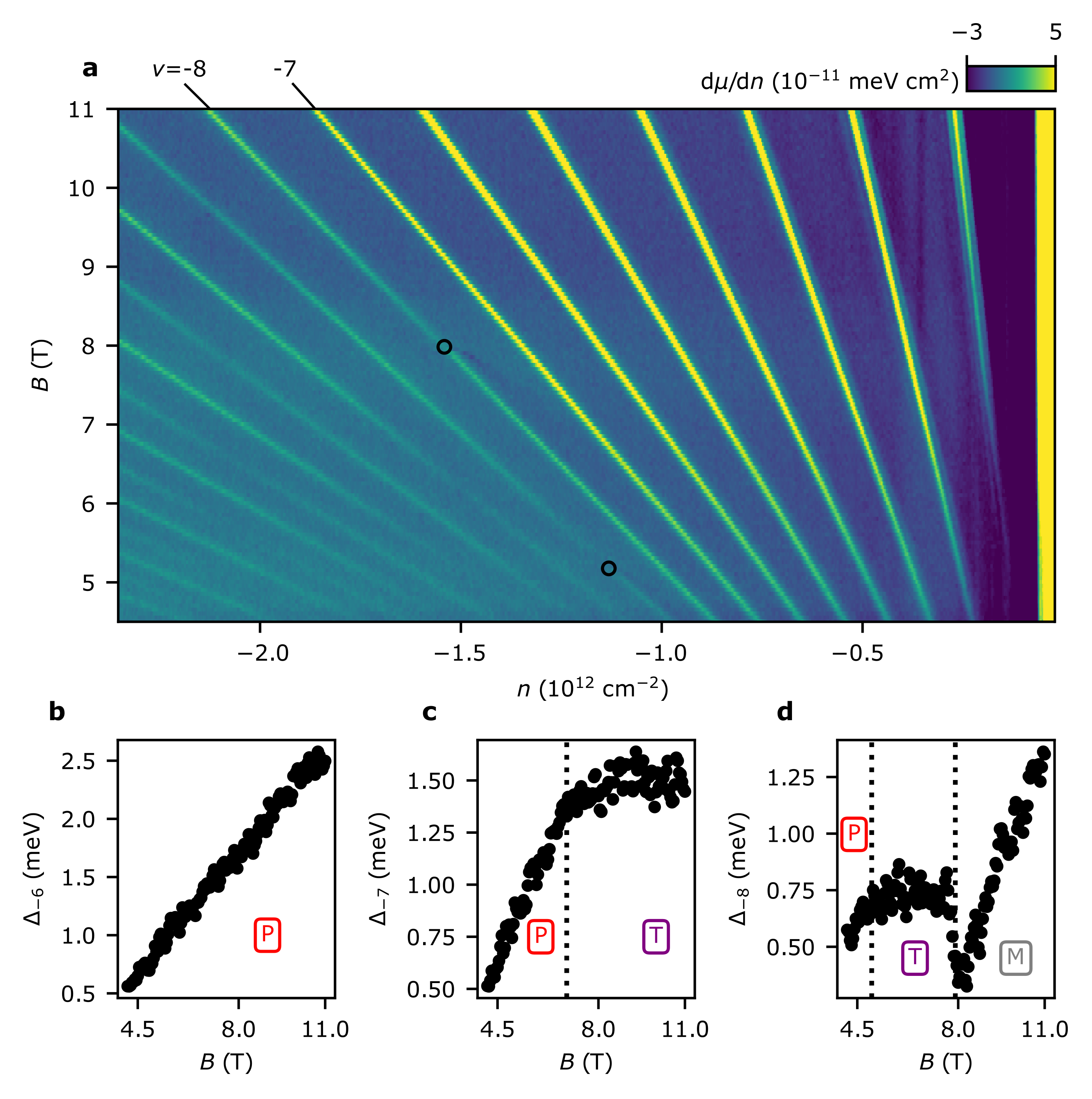}
    \caption{\textbf{LLs in a Bernal bilayer flake of WSe$_2$.} \textbf{a}, Landau fan of d$\mu$/d$n$ as a function of $n$ and $B$ in a Bernal bilayer flake of WSe$_2$. Similar phase transitions occur but at different densities / magnetic fields. \textbf{b-d}, LL gap sizes for the $\nu = -6$, $\nu = -7$, and $\nu = -8$ gaps. Both $\nu = -7$ and $\nu = -8$ gaps show plateaus as a function of $B$ preceding magnetic phase transitions, similar to the behavior in monolayers.} 
    \label{fig:BLLLs}
\end{figure*}

In Fig. \ref{fig:BLLLs}, we present similar data measured in a separate device (Sample D) on a naturally-stacked ($2H$, or Bernal-stacked) bilayer WSe$_2$ flake. Given the geometry of our sample, which does not have a top gate, hole doping the bilayer intrinsically leads to a nonzero effective displacement field. Because of the lack of interlayer coupling in bilayer WSe$_2$, this measurement is likely to result in full layer polarization, filling LLs in the bottom of the two layers while the second layer remains in the semiconducting band gap \cite{shi_bilayer_2022,shih_spin-selective_2023}. This bilayer sample therefore acts effectively like a monolayer. Indeed, we observe qualitatively similar gap dependence on magnetic field, including sharp drops at the onset of the ``mixed" regime and negative compressibility, as well as plateaus in the gap sizes (Fig. \ref{fig:BLLLs}c-d).  In line with previous reports, the positions of the transitions quantitatively change, as fewer polarized LLs are preferentially populated at a given density compared to the monolayer case \cite{shi_bilayer_2022}. This could stem from a difference in the dielectric environment felt by the holes in the lower layer. Quantitatively, we observe a similar effective mass for the polarized LLs $m^* = 0.38 \pm 0.01 \mZero$, within the range of monolayer samples that we measured.

\clearpage

\section{Band structure details}\label{sec:BandStructure}

We want to relate the band parameters to the parameters that appear in the interacting description of our system. First principle calculations \cite{xiao_coupled_2012, kormanyos_kp_2015} have shown that WSe$_2$ is will described by a $\vec{k}\cdot\vec{p}$ Hamiltonian of the form: 
\begin{equation}\label{eq:k.p}
    h(\vec{q}) = \frac{\Delta}{2}\sigma_z + \lambda\tau s \frac{1-\sigma_z}{2} + v \vec{q}\cdot \vec{\sigma}_{\tau} + \frac{\vec{q}^2}{4\mZero}\left(\alpha+\beta \sigma_z\right) 
\end{equation}
where $\vec{q}$ is the momentum of the electron measured with respect to the band minima ($K$ for $\tau=+1$ and $-K$ for $\tau=-1$), $\vec{\sigma}_\tau = (\tau \sigma_x,\sigma_y)$ is a vector of Pauli matrices acting in the conduction/valence basis, $\mZero$ is the electron mass in vacuum, $v$ is a velocity, $\Delta$ is a mass gap without spin-orbit coupling (SOC), and $\lambda$ is a SOC gap. Here $\tau=\pm 1$ is a valley label and $s=\pm 1$ is a spin label ($s=+1$ for spin up and $s=-1$ for spin down). Note that the parameters $\alpha$ and $\beta$ control the particle-hole asymmetry. 


We diagonalize the Hamiltonian in Eq.~\ref{eq:k.p} in the presence of a magnetic field $B$ perpendicular to the sample. More precisely, we use a Peierls substitution $\vec{q}\to \vec{q} - e \vec{A}$ where $\vec{A}$ is the gauge potential ($\vec{\nabla}\times\vec{A} = B \hat{\mathbf{z}}$, and we add the following terms linear in $B$: $\frac{g_s}{2} \mu_{B} sB$ (Zeeman energy of electrons in vacuum), and  $g_v m_v \mu_B\tau (1-\sigma_z)/2$ (magnetic moment due to the orbital character of the conduction and valence band near the $\pm K$ points). We take the values $g_s\approx 2$, $g_v\approx 1$ and $m_v\approx 2$ used in Ref.~\cite{gustafsson_ambipolar_2018}.

Diagonalizing the Hamiltonian, we obtain 4 different sets of LLs labeled by their (Dirac) Landau level index $N_{D}= 0,\pm 1,\pm 2, \dots $, spin $s$ and valley $\tau$. The LLs coming from the upper valence band at $B=0$ have $s=\tau$. Due to the different Berry phase for each valley, these LLs have indices $N_D=0,-1,-2,-3,\dots$ for $s=+1(\uparrow)$; and $N_D=-1,-2,-3,\dots$ for $s=-1(\downarrow)$. 

Following the main text, we label the hole Landau levels by positive integers $N \equiv - N_{D}$. Due to the large gap between the valence and conduction band ($\Delta -\lambda$), the Landau wave-functions for the LL with index $N$ are well approximated by $[0,\phi_{N-(1-s)/2}]^{\top}$ where $\phi_{N}$ are the standard wave-functions of ($N+1$) lowest LL of electrons with a quadratic dispersion. This result can be derived by inspecting the exact eigenfunction for the Landau levels.

We extract the bare cyclotron and bare Zeeman energies used in Sec.~\ref{sec:TheoryGaps} as 
\begin{equation}\label{eq:bandEnergies}
    \begin{split}
        \frac{\Ecb}{B} &= \lim_{B\to 0}\frac{\varepsilon_{0++}(B) - \varepsilon_{1,++}(B)}{B} \\ &= \left(\hbar  e\right) \left[ \frac{2v^2}{\Delta-\lambda} +\frac{\beta-\alpha}{2\mZero}\right]\\
        \frac{\EZb }{B} &= \lim_{B\to 0}\frac{\varepsilon_{0++}(B) - \varepsilon_{1,--}(B)}{B}\\
        &= \frac{\hbar e}{\mZero} \left[\frac{g_s}{2}+g_v m_v + \frac{\alpha-\beta}{2}\right] + \frac{\Ecb}{B}.
    \end{split}
\end{equation}
where $\varepsilon_{N s \tau}$ is the energy of the LL with index $N$, spin label $s$, and valley label $\tau$.

The above expression for $\Ecb$ is equal to $\hbar e B /m$ where $m$ is the ``band mass" obtained by matching the quadratic term in $\vec{q}$ of the eigenvalues of the Hamiltonian in Eq.~\ref{eq:k.p}. Therefore, we can use the effective mass measured using ARPES to determine $\Ecb$ and we do not require the use of first-principle parameters. However, due to the lack of measurements of the Zeeman energy in a weakly interacting regime, we are forced to use first-principle parameters for this purpose. Different band structure parameters give different values of $\EZb / \Ecb$. To be explicit, where needed we will use the parameters from Ref.~\cite{xiao_coupled_2012}, for which we find 
\begin{equation}\label{eq:DefinitionChi0}
    \chi_0 \equiv \EZb/\Ecb = 2.05,
\end{equation} 
where we have defined the useful parameter $\chi_0 $ as the ratio of the band Zeeman energy over the band cyclotron energy. The parameters from Ref.~\cite{xiao_coupled_2012} neglect the particle-hole asymmetry ($\alpha=\beta =0 $). Using the values of $\alpha-\beta$ from Ref.~\cite{kormanyos_kp_2015} and the ARPES value for the effective mass, we find $\EZb/\Ecb = 1.94$. We can extract a $g$-factor via $\frac{g}{2} = \frac{\mZero\EZb}{\hbar e B }$ and we use the value $\EZb/\Ecb = 2.05$ our theoretical calculations (except in Fig. \ref{fig:QMCComparison} and Fig. 1c in the main text where we show a range of values for $\chi_0$ around 2). 

To summarize, we have found that in the regime of interest, the valence band LLs of WSe$_2$ are the same as those of a conventional two-dimensional hole gas. In particular, the orbital wave-functions for the LLs with spin up and index $N$ are the same as in a conventional hole gas. On the other hand, LLs of spin-down holes with index $N$ have the same orbital wave-function as a conventional hole gas with index $N-1$ but only the LLs with $N\geq 1$ are present. Without loss of generality, hereafter we apply a particle-hole transformation to make analogy to the more familiar case of carriers that are electrons. 

\section{Theoretical calculation of excitation gaps at integer fillings}\label{sec:TheoryGaps}

\subsection{Setup}
As argued in Sec.~\ref{sec:BandStructure}, the single-particle physics of WSe$_2$ is well-described, after a particle-hole transformation, by a model of electrons with a conventional quadratic dispersion. We now proceed to introduce interactions. 

In the absence of a magnetic field, the system is described by 
\begin{equation}
    H = \sum_{j} \frac{\vec{p}_j^2}{2m} + \sum_{j<j'} u(\vec{r}_j -\vec{r}_{j'}),
\end{equation}
where $\vec{r}_i$ and $\vec{p}_i$ are the position and momentum of the $i$-th electron. $u(\vec{r}_j -\vec{r}_{j'})$ is the inter-particle potential, which is discussed below. 

In the presence of a magnetic field perpendicular to the sample, two modifications are needed. First, we send $\vec{p}_j\to \vec{p}_j +\vec{A}_j(\vec{x}_j)$, were $\vec{A}_j$ is the vector potential that satisfies $\vec\nabla_{j}\times \vec{A}_{k} =\delta_{jk} B \hat{\bf z}$. Second, we need to add a Zeeman energy ($ - \frac{g}{2}\mu_{B} B\sigma $) for each electron, where $\mu_{B} = \frac{e\hbar}{2m_{\text{e}}}$ is the Bohr magneton. Here $g$ is the``band g-factor" extracted from the band structure as specified in Eq.~\ref{eq:bandEnergies}. This $g$ already takes into account the orbital and spin Zeeman couplings present in the band Hamiltonian at finite $B$ field. Note that the value of $g$ is different for a magnetic field with an in-plane component. Previous experiments suggest that in-plane $g$-factor is small \cite{movva_density-dependent_2017}. 

Putting things together, the Hamiltonian in the presence of magnetic fields is
\begin{equation}
    H = \sum_{j} \frac{(\vec{p}_j - e \vec{A}_j)^2}{2m}  - \frac{g}{2} \mu_{B} B \sigma_j + \sum_{j<j'} u(\vec{r}_j -\vec{r}_{j'}).
\end{equation}

\begin{figure*}[h]
    \renewcommand{\thefigure}{S\arabic{figure}}
    \centering
    \includegraphics[width=0.6\textwidth]{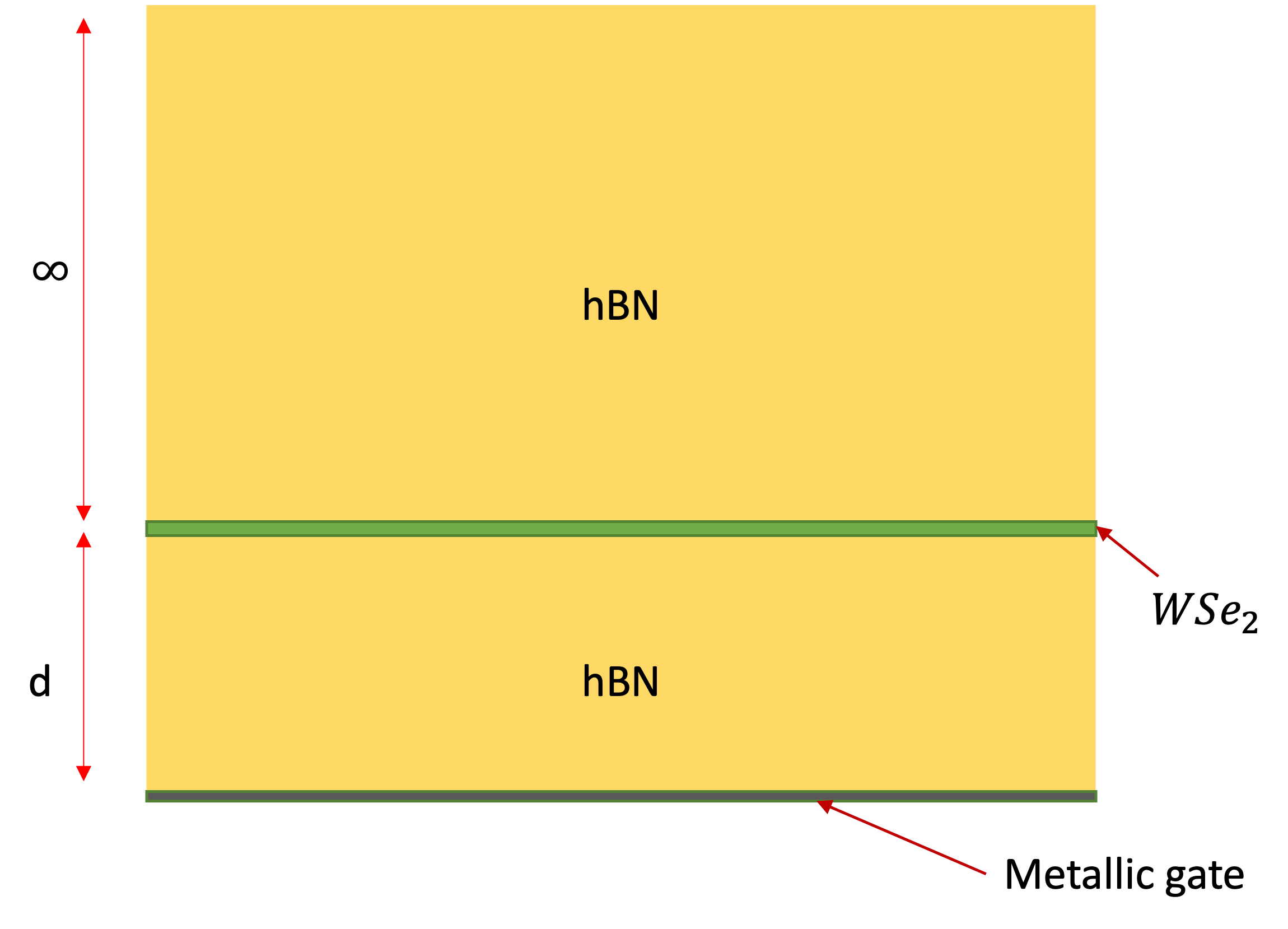}
\caption{\textbf{Dielectric enviroment used in the theoretical calculation}. Monolayer WSe$_2$ is encapsulated by hexagonal boron nitride (hBN) and is separated by a distance $d$ from a metallic back gate.} 
    \label{fig:DielectricEnviroment}
\end{figure*}

We take the potential to be $u(\vec{r}) = \int\dd{\vec{q}} e^{i \vec{q}\cdot\vec{r}} \tilde{u}(\vec{q})$ with 
\begin{equation}
    \tilde{u}(\vec{q}) = \frac{e^2}{4\pi \epsilon \abs{\vec{q}}}\frac{1-e^{-\xi \abs{\vec{q}}}}{1+r_0 \abs{\vec{q}}}
\end{equation}
where the parameters $\xi,\epsilon, r_0$ depend on the sample. $\xi$ and $\epsilon$ are obtained solving the electrostatic problem of a test charge placed in the WSe$_2$ layer of the dielectric environment in Fig.~\ref{fig:DielectricEnviroment}.

We find that $\epsilon = \sqrt{\epsilon_{\text{hBN},\parallel} \epsilon_{\text{hBN},\perp} }$ and $\xi = 2\sqrt{\frac{\epsilon_{\text{hBN},\parallel} }{\epsilon_{\text{hBN},\perp} }} d$, where $d$ is the distance from the sample to the back gate, $\epsilon_{\text{hBN},\parallel}$ is the in-plane dielectric constant of hBN, and $\epsilon_{\text{hBN},\perp}$ is the out-of-plane dielectric constant of hBN. We use the values of $\epsilon_{\text{hBN},\parallel}=6.93$ and $\epsilon_{\text{hBN},\perp}=3.76$ from Ref.~\cite{laturia2018dielectric}. The distance $d$ is set by the separation from the back gate, i.e. the ``Bottom hBN thickness" reported in Tab.~\ref{tab:Tab1}. In particular, we use $\epsilon = 5.1$ and, for Sample A,  $\xi = 120\ \text{nm}$.

The Keldysh length $r_0$ is a parameter that captures the fact that the wave-function has a finite extend in the out-of-plane direction, as well as the dielectric constant mismatch between the sample and the environment \cite{keldysh1979coulomb}. Previous studies have estimate this to be of the order of 1~nm \cite{ Berkelbach2013Dielectric,Belgium2018Dielectric-Excitons-TMDs}. However, we find that a value of 0.5~nm gives better agreement with the gaps measured in experiment. 

\subsection{Estimation of gaps}\label{sec:RPA_expV}

The experimentally measured gaps are equivalent to the difference between the energy of creating a particle and a hole. Therefore, we need to evaluate the the energies of creating particles and holes on top of a reference state. The reference states we consider have fully filled LLs and are labeled by the number of filled LLs for each spin $(\nu_{\uparrow},\nu_{\downarrow})$, where the total filling factor is given by $\nu = \nu_{\uparrow} + \nu_{\downarrow}$. We evaluate these energies within the Hartree-Fock approximation but with the potential screened within the random phase approximation (RPA). Using this screened potential takes into account some effects of LL mixing as it captures virtual transitions between filled and empty LLs. 

The screened interaction potential within the RPA is given by 
\begin{equation}\label{eq:ScreenedPotential}
    \tilde{u}_{\text{eff}}(\vec{q}) =  \frac{\tilde{u}(\vec{q})}{1-\tilde{u}(\vec{q})[\Pi_{\nu_{\ua}}(\omega = 0 , \vec{q})+\Pi_{\nu_{\da}}(\omega = 0 , \vec{q})]},
\end{equation}
where $\Pi_{\nu}(\omega,\vec{q})$ is the Lindhard function of the lowest $\nu$ LLs of a spin-polarized conventional electron gas (see e.g. Chapter 10 of Ref.~\cite{giuliani2005quantum} for expressions for $\Pi_{\nu}$). 

Within Hartree-Fock, the energy of a particle in the $M^{th}$-lowest spin $\sigma$ LL is
\begin{equation}\label{eq:energyParticle}
    \varepsilon_{M,\sigma} = \Ecb\left( M + \frac{1}{2} \right) - \sigma \frac{\EZb}{2}  + \varepsilon^{\text{(d)}}+ \varepsilon_{M,\sigma}^{\text{(x)}}- \mu.
\end{equation}
where $\mu$ is the chemical potential, $\varepsilon_{M,\sigma}^{\text{(x)}}$ is the exchange (Fock) contribution and $\varepsilon^{\text{(d)}}$ is the direct (Hartree) contribution that is independent of $M$ and $\sigma$. 
We evaluated $\varepsilon_{M,\sigma}^{\text{(x)}}$ using the expressions in Ref.~\cite{Kallin1984ExcitationsFilledLLs2DEG} with the bare potential replaced by the screened one:
\begin{equation}
    \varepsilon_{M,\sigma}^{\text{(x)}} = 
    -\sum_{N'=0}^{\nu_{\sigma}-1} \int \frac{d^2{\vec{r}}}{2\pi\lB^2} u_{\text{eff}}(\vec{r}) L_{M-1}(r^2\lB^2/2) L_{N'}(r^2\lB^2/2)e^{- r^2\lB^2/2}
\end{equation}
where $L_{n}(x)$ are the Laguerre polynomials of order $n$ and ${u}_{\text{eff}}(\vec{r})$ is the inverse Fourier transform of $\tilde{u}_{\text{eff}}(\vec{q})$ in Eq.~\ref{eq:ScreenedPotential} and $\lB = \sqrt{\abs{\frac{\hbar}{eB}}}$ is the magnetic length.

\begin{figure*}[h]
    \renewcommand{\thefigure}{S\arabic{figure}}
    \centering
    \includegraphics[scale=1.0]{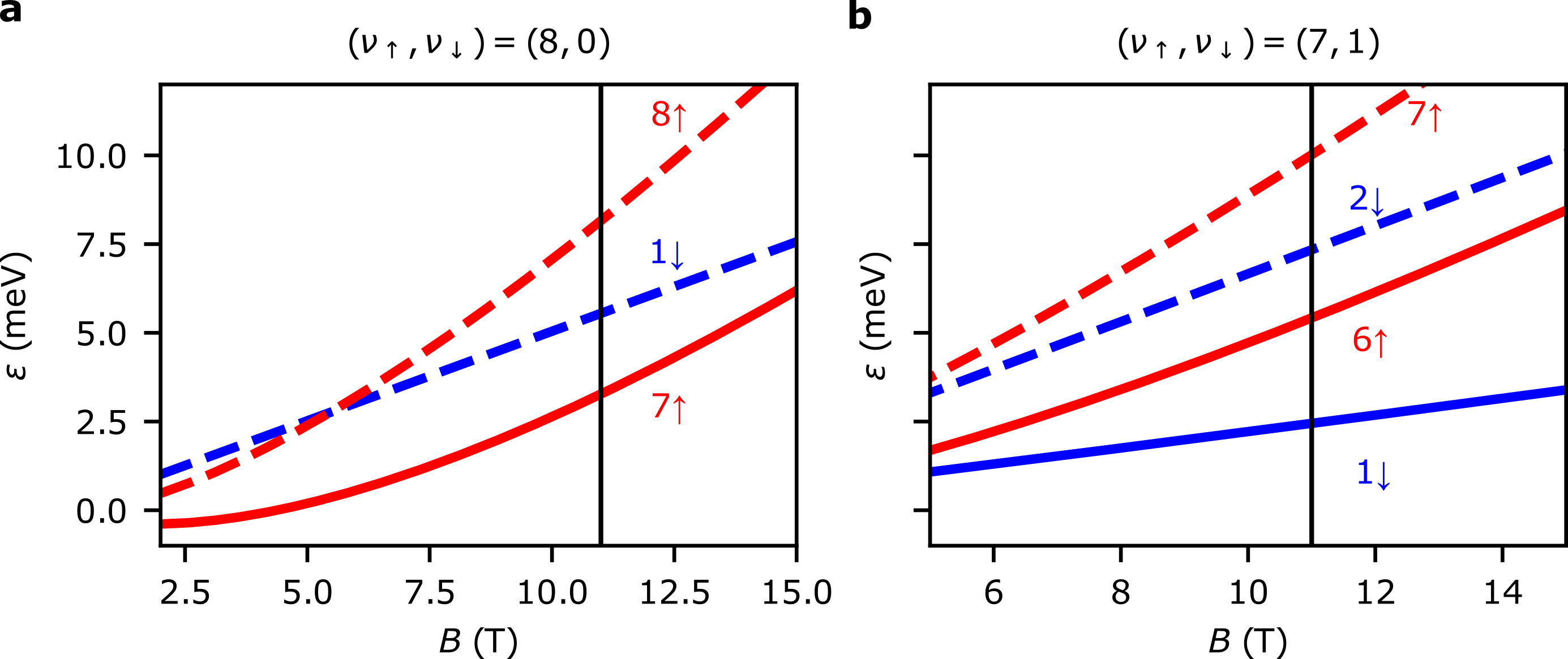}
    \caption{\textbf{Energies of various LLs calculated within the HF approximation using RPA-screened interactions}. \textbf{a-b}, Calculated energy to add a particle to (remove a hole from) a given empty (filled) LL as a function of $B$ for two different ground states at filling $\nu = -8$: the fully polarized state (\textbf{a}) and the state with one spin-minority LL fully populated (\textbf{b}). Spin majority (minority) LLs are colored in red (blue), while empty (filled) LLs are dashed (solid). Numbers indicate the orbital index $N$. The black vertical lines correspond to the first-order transition at $B_{\rm{C}}$. The gaps were plotted using the parameters from Sec.~\ref{sec:TheoryGaps} for for Sample A ($m = 0.35 m_{e}$, $\epsilon = 5.10$, $\chi_0 = 2.05$ $r_0= 0.5\ \text{nm}$ and $\xi = 120\ \text{nm}$).}
    \label{fig:MixedLLEnergies}
\end{figure*}

Figure~\ref{fig:MixedLLEnergies} shows the energy to create (remove) a particle (hole) in the LLs close to the Fermi level for the states $(\nu_{\ua},\nu_{\da})= (8,0)$ and $(\nu_{\ua},\nu_{\da})= (7,1)$. The vertical black line marks the magnetic field at which we observe the first-order transition in our experiments. These energies were calculated using Eq.~\ref{eq:energyParticle} with a Keldysh length of $r_0=0.5$~nm. We plotted the energy of the LLs relative to $\varepsilon^{\text{(d)}}-\mu$. Note that the energy of the LLs for the $(\nu_{\ua},\nu_{\da}) = (8,0)$ shown in Fig.~\ref{fig:MixedLLEnergies}a tell us that at small $B$, the lowest excitations correspond to particles and holes with spin up, but at a field $B_T\approx 5.5$ T, the order of states switches such that for larger $B$ the spin-down particle excitation has lower energy than the spin-up particle excitation. This is consistent with the observation in the main text of a transition between the ``P" and the ``T" regimes.

According to our experiments and the discussion below in Sec.~\ref{sec:VMC}, 
we expect a first-order transition from a ground state $(\nu_{\ua},\nu_{\da}) = (8,0)$ to $(7,1)$ above the critical field $B_{\rm C} \approx 11$ T. Thus the gaps for $B \gtrsim B_{\rm C}$ should be calculated using the $(\nu_{\ua},\nu_{\da})=(7,1)$ state. Fig.~\ref{fig:MixedLLEnergies}b shows that for this ground state in the mixed regime, the lowest energy hole is predicted to be spin up and the lowest energy particle is predicted to be spin down. Additionally, the precise ordering of relevant states in the mixed regime depends on the overall filling factor $\nu$. We had expected the energy of the LLs to be such that the lowest energy hole has spin down while the lowest energy particle has spin up (see Fig.~2l in the main text). We believe the reason for the discrepancy is that our approximation does not capture correlation effects, which may be more important once we have we have a finite density of filled states with both spin up and down, as is the case throughout the mixed LL regime. We expect considering such effects will push the spin-up (spin-down) LLs down (up) in energy. Because of this nuance, we do not plot the theoretically predicted gaps beyond $B_{\rm C}$ in Fig.~2f-j in the main text, and we leave the detailed behavior in this regime to a future calculation. 

\begin{figure*}[h]
    \renewcommand{\thefigure}{S\arabic{figure}}
    \centering
    \includegraphics[scale=1.0]{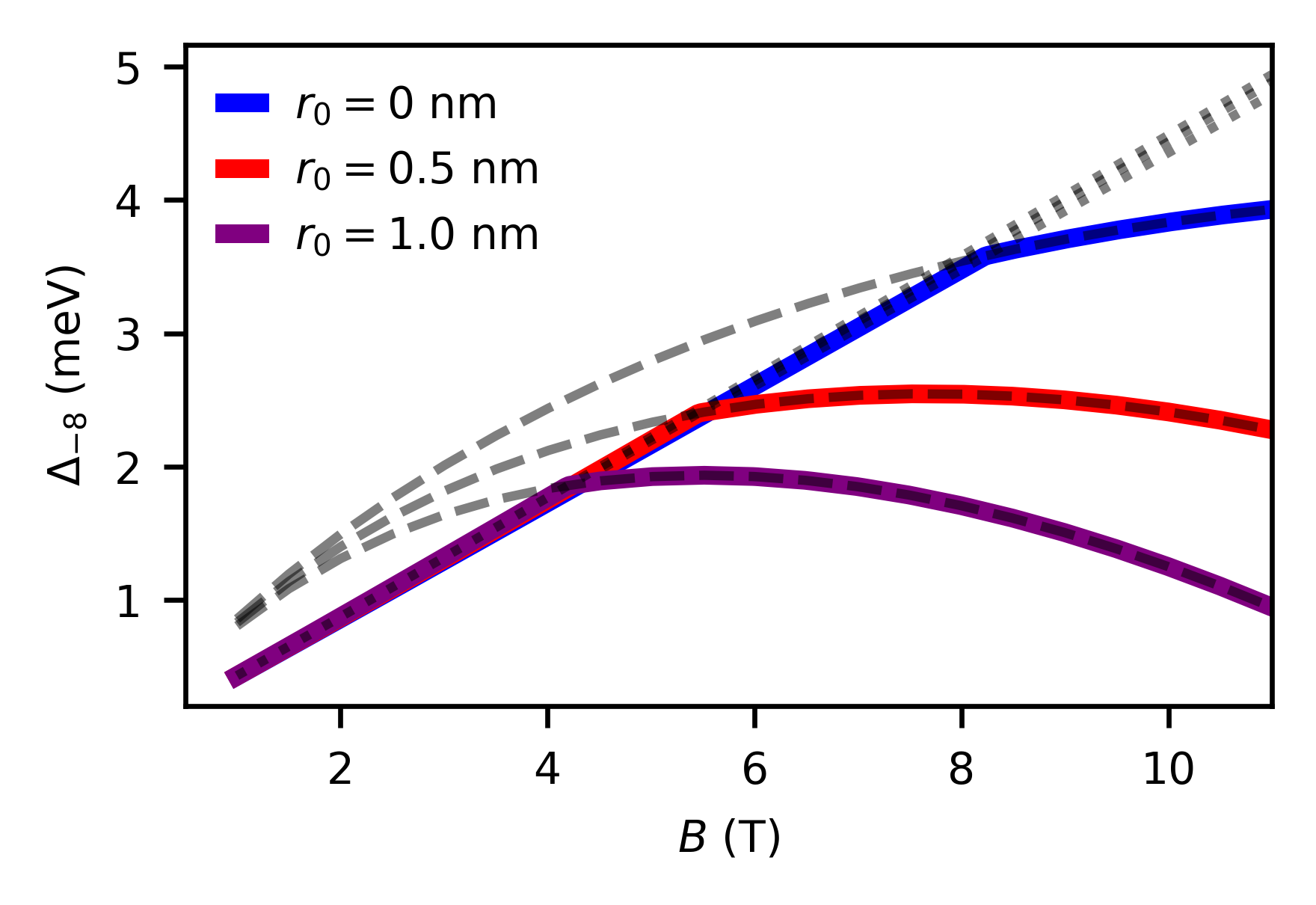}
    \caption{\textbf{Calculated $\nu=-8$ gap for different Keldysh parameters.} Calculated LL gap at $\nu = -8$ corresponding to different choices of the Keldysh parameter $r_0$. We also show the calculated $\Delta_{\cyc}$ (dotted lines) and $\Delta_{\text{sf}}$ (dashed lines) for each gap; the experimentally relevant gap (thick colored lines) is the minimum of these two energies. The main effect of increasing the Keldysh parameter is to decrease the critical magnetic field $B_T$ above which the spin-flip excitation is preferred. These gaps were calculated using the parameters in Sec.~\ref{sec:TheoryGaps} for Sample A ($m = 0.35 m_{e}$, $\epsilon = 5.10$, $\chi_0 = 2.05$ and $\xi = 120\ \text{nm}$). }
    \label{fig:Keldysh}
\end{figure*}

We now focus on the gaps in the low field regime where $(\nu_{\uparrow},\nu_{\downarrow} )= (\nu,0)$. In this case, the hole must have spin up but the particle can have spin up or down. Thus, we can define a gap for each spin orientation of the particle: 1) a cyclotron gap $\Delta_{\cyc} = \varepsilon_{\nu,\uparrow} - \varepsilon_{\nu-1,\uparrow}$; and 2) a spin-flip gap $\Delta_{\text{sf}} = \varepsilon_{0,\downarrow} - \varepsilon_{\nu-1,\uparrow}$. The gap to create a particle-hole excitaiton is the smaller of the two: $\Delta_{\text{ph}} = \min(\Delta_{\cyc}, \Delta_{\text{sf}})$. Figure 2 in the main text and Fig. \ref{fig:Keldysh} show $\Delta_{\cyc}$ and $\Delta_{\text{sf}}$ in dotted and dashed lines, respectively, and the corresponding $\Delta_{\text{ph}}$ (measured in Fig.~2a-e or predicted in Fig.~\ref{fig:Keldysh}) with solid lines. 

We note that the cyclotron and spin-flip gaps can be written as a sum of two contributions: $\Delta_{\text{j}} = \Delta_{\text{j}}^{(\text{0})} + \Delta_{\text{j}}^{\text{(x)}}$, where {j} = {cyc} or {sf}. The ``bare gap" $\Delta_{\text{j}}^{(\text{0})}$, is defined to be the difference of the non-interacting energy between a particle and hole. An ``exchange gap" $\Delta_{\text{j}}^{(\text{x})}$ is defined to be the difference of the exchange energy between the particle and hole. As $B$ increases, interaction effects are less important so $\Delta^{\text{(0)}}_{\text{j}}$ dominates. In contrast, at small $B$, interaction effects become more important so $\Delta^{\text{(x)}}_{\text{j}}$ dominates.

We now proceed to determine the $B$-dependence of the cyclotron and spin-flip gaps by looking at each of their contributions. The bare cyclotron gap is simply the bare cyclotron energy $\Delta^{\text{(0)}}_{\cyc} = \Ecb$ which is linear in $B$ and positive. The exchange gap $\Delta^{\text{(x)}}_{\cyc}$ is also positive and an increasing function of $B$ because the hole can exchange more effectively than the particle with the filled LLs because their form-factor is closer to that of the filled LLs. However, $\Delta^{\text{(x)}}_{\cyc}$ is small when $\nu$ is large because the form-factor does not change much in this regime. 

For the spin flip gap, the bare contribution is $\Delta^{\text{(0)}}_{\text{sf}} = (1-\nu)\Ecb + \EZb 
$ which is negative for large enough $\nu$.  The exchange contribution $\Delta^{\text{(x)}}_{\text{sf}}$ is positive and large as no cancellation happens because the spin-down particle doesn't have filled states to exchange with. Additionally, $\Delta^{\text{(x)}}_{\text{sf}}$ is an increasing function of $B$. To see this, note that increasing $B$ implies increasing $n$ too because we are at fixed filling $\nu$. Then, it is natural that $\Delta^{\text{(x)}}_{\text{sf}}$ increases because now the spin-up hole has more electrons to exchange with. 

At low $B$ field, exchange dominates and $\Delta^{}_{\text{sf}}\approx \Delta^{\text{(x)}}_{\text{sf}}$, which implies that the gap is an increasing function of $B$ for small $B$. However, as $B$ increases, the bare gap starts to dominate, thus making $\Delta^{}_{\text{sf}}$ a decreasing function of $B$. Therefore, $\Delta^{}_{\text{sf}}$ should have a maximum as a function of $B$. The apparent flatness of the experimentally measured gaps can be explained by a small curvature of $\Delta^{\text{(0)}}_{\text{sf}}$ around this maximum (which is not guaranteed a priori), and that this flat region occurs in a region where $\Delta^{\text{(0)}}_{\text{sf}} < \Delta^{\text{(0)}}_{\text{cyc}}$.

\section{Estimation of $B_{\rm{C}}$ using Quantum Monte Carlo}\label{sec:VMC}

In this section, we aim to give, for each $\nu$, an estimate of the critical magnetic field, $B_{\rm C}(\nu)$, at which the behavior of the gap changes from the ``transition" regime to the ``mixed" regime. We consider a pure Coulomb interaction $u(\vec{r}) = \frac{e^2}{4\pi \epsilon \abs{\vec{r}}}$ so that the electron gas is now described by the band mass $m$, the dielectric constant $\epsilon$, the bare $g$ factor and its density $n$. Following the two-dimensional electron gas literature, we measure energies in units of the effective Hartree $\text{Ha}_* = m \left(\frac{e^2}{4\pi \epsilon \hbar}\right)^2$, and denote by $\nu_{\sigma}\geq 0$ the number of electrons with spin $\sigma$ per unit flux. 

To estimate $B_{\rm C}$, we assume that the transition occurs when the fully-polarized ground state ($\nu_{\ua} = \abs{\nu}$,$\nu_{\da}=0$) becomes degenerate with the partially polarized ground state ($\nu_{\ua} = \abs{\nu}-1$, $\nu_{\da}=1$). Let $\tilde{\e}(\nu_{\ua},\nu_{\da},B)$ be the energy per electron of the state with $\nu_{\ua}$ spin-$\ua$ electrons and $\nu_{\da}$ spin-$\da$ electrons per unit flux in the presence of a out-of-plane magnetic field $B$. Then $B_{\rm C}$ for the $\nu$ state satisfies
\begin{equation}
    \tilde{\e}(\abs{\nu},0, B_{\rm C}) = \tilde{\e}(\abs{\nu}-1,1, B_{\rm C}).
\end{equation}

For future convenience, let $\e(\rs,\zeta) $ be the energy per electron of the conventional electron gas at zero magnetic field as a function of the Wigner-Seitz radius $\rs = \frac{1}{\sqrt{\pi (n_{\ua} + n_{\da})}\aB}$ and polarization $\zeta = \frac{n_{\ua} - n_{\da}}{n_{\ua} + n_{\da}}$, where $\aB = \frac{\hbar^2}{m} \frac{4\pi\epsilon}{e^2}$ is the effective Bohr radius, and $n_{\sigma}$ is density per area of electrons with spin $\sigma = \ua,\da$.

Within the Hartree-Fock approximation and the Random Phase approximation, $\tilde{\e}(\nu_{\ua},\nu_{\da},B)$ is well approximated by $\e(\rs, \zeta) - \frac{\chi_0}{\rs^2}\frac{\zeta}{\nu_{\ua} + \nu_{\da}}$ where $\chi_0 = \frac{g m }{2\mZero}$ and the parameters $\rs, \zeta$ are evaluated using $n_{\sigma} = \nu_{\sigma} \frac{e B}{h }$ \cite{giuliani2005quantum}. 

Both of the above approximations overestimate interaction effects in $\e(\rs,\zeta)$ such that they predict a transition from the paramagnetic state ($\zeta=0$) to the polarized state ($\zeta=1$) at a smaller $\rs $ than more accurate methods. By using the Quantum Monte Carlo (QMC) method to calculate the energies per electron, we can predict a more reasonable critical $\rs$ for the above transition \cite{attaccalite_correlation_2002, attaccalite_erratum_2003}. Let $\e_{\QMC}(\rs,\zeta)$ be the energy per electron calculated using the parametrization of Refs.~\cite{attaccalite_correlation_2002, attaccalite_erratum_2003}.

From the above two observations, we approximate the energy per electron as 
\begin{equation}\label{eq:energyAnsatz}
    \tilde{\e}(\nu_{\ua},\nu_{\da},B) \approx \e_{\QMC}(\rs,\zeta)-\frac{ \chi_0 }{\rs^2}\frac{\zeta}{\nu_{\ua}+\nu_{\da}},
\end{equation}
where again $\rs$ and $\zeta$ are calculated using $n_{\sigma}=\nu_{\sigma} \frac{eB}{h}$ and $\chi_0 = \frac{gm}{2\mZero}$. 

\begin{figure*}[h]
    \renewcommand{\thefigure}{S\arabic{figure}}
    \centering
    \includegraphics{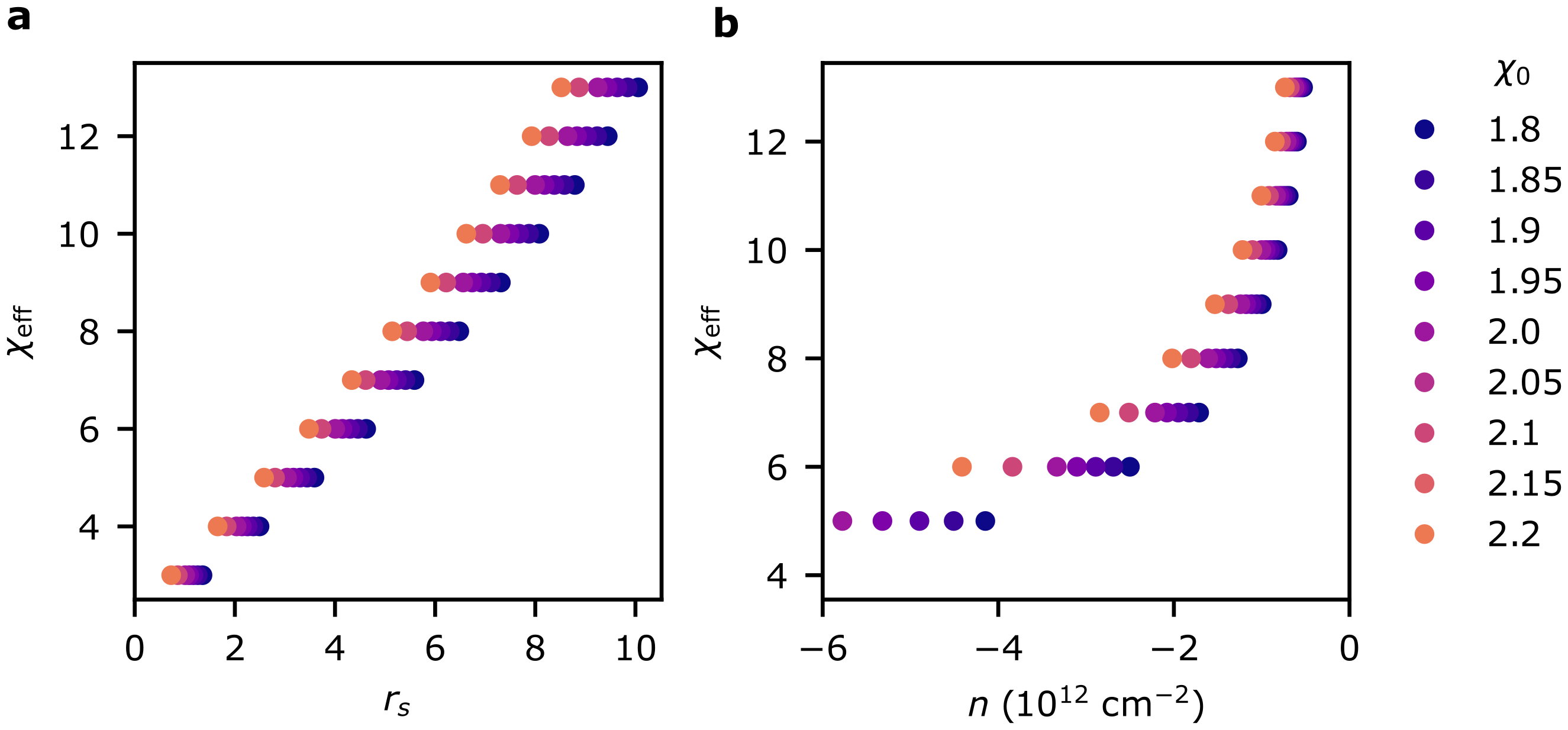}
    \caption{\textbf{Spin susceptibility estimated via quantum Monte Carlo calculations}. \textbf{a-b}, Calculated effective spin susceptibility $\chi_{\text{eff}}$ for a two-dimensional electron gas with appropriate parameters for WSe$_2$ as a function of $r_s$ (\textbf{a}) and carrier density (\textbf{b}) for various values of the single-particle Zeeman coupling $\chi_0$. } 
    \label{fig:QMCComparison}
\end{figure*}

We determine an effective magnetic susceptibility $\chi_{\text{eff}}$ as a function of $\rs$ following the experimental definition of $g^*$ in Ref.~\cite{shi_odd-_2020}. $\chi_{\text{eff}}$ is equal to $(\abs{\nu}-1)$ at $B=B_{\rm C}(\nu)$ or equivalently at $\rs = \frac{1}{\sqrt{\pi \abs{n}} \aB}$ with $n= \nu \frac{eB_{\rm C}}{h}$. In Fig.~\ref{fig:QMCComparison} we show $\chi_{\text{eff}}$ in the range relevant for the experiment as a function of $\rs$ and $n$ using the parameters $\epsilon = 5.10$ and $m = 0.35 \mZero$. We used the values of $\chi_{\text{eff}}$ between $\chi_0 = 2.2$ and $\chi_{0} = 1.8$ to plot the gray region in Fig. 1c in the main text.

\section{Phenomenological behavior of Landau levels near transitions}\label{Sec:LLPhenomenology}

Here, we expand on the discussion in the main text and detail additional phenomenological aspects of the LLs in this system, accounting for finite disorder and temperature. Using a model that incorporates appropriate LL broadening, we can understand both the slight density shift in the incompressible LL peak at the first order phase transitions (Fig. 3 in the main text) as well as the decreased negative compressibility at low temperatures.   

At the first-order phase transitions in the LL gap size, where the gap size sharply drops and hysteresis is observed, there is also a slight shift in the filling factor at which the incompressible peak is observed. This is most clearly seen in Fig. 3b-c in the main text, where the incompressible peak jumps slightly left, to higher hole doping, above the critical field $B_{\rm{C}}$. One possible cause of a shift would be a change in the quantum capacitance of the WSe$_2$ monolayer, which would lead to a change in conversion between the DC voltage applied to the back gate and the density induced in the monolayer \cite{eisenstein_compressibility_1994,shi_odd-_2020}. However, we rule out this explanation because we explicitly measure d$\mu$/d$n$, which is proportional to the inverse of the quantum capacitance. The (average) change in quantum capacitance at partial LL filling above and below the phase transition would need to correspond to a d$\mu$/d$n$ of roughly $10^{-10}$ meV cm$^{2}$ to account for the shift of the incompressible peak if the peak were to occur at fixed LL filling. Because we observe no such shift (the noise floor of d$\mu$/d$n$ is an order of magnitude lower and there is no obvious change in background across the transition), we do not believe that this is a possible explanation for the data.

An alternative scenario is that the incompressible peak at the transition is not perfectly centered at integer filling factor. This can arise if disorder broadens the LL density of states in a manner that varies as a function of the LL index, as we detail below.

\begin{figure*}[h]
    \renewcommand{\thefigure}{S\arabic{figure}}
    \centering
    \includegraphics{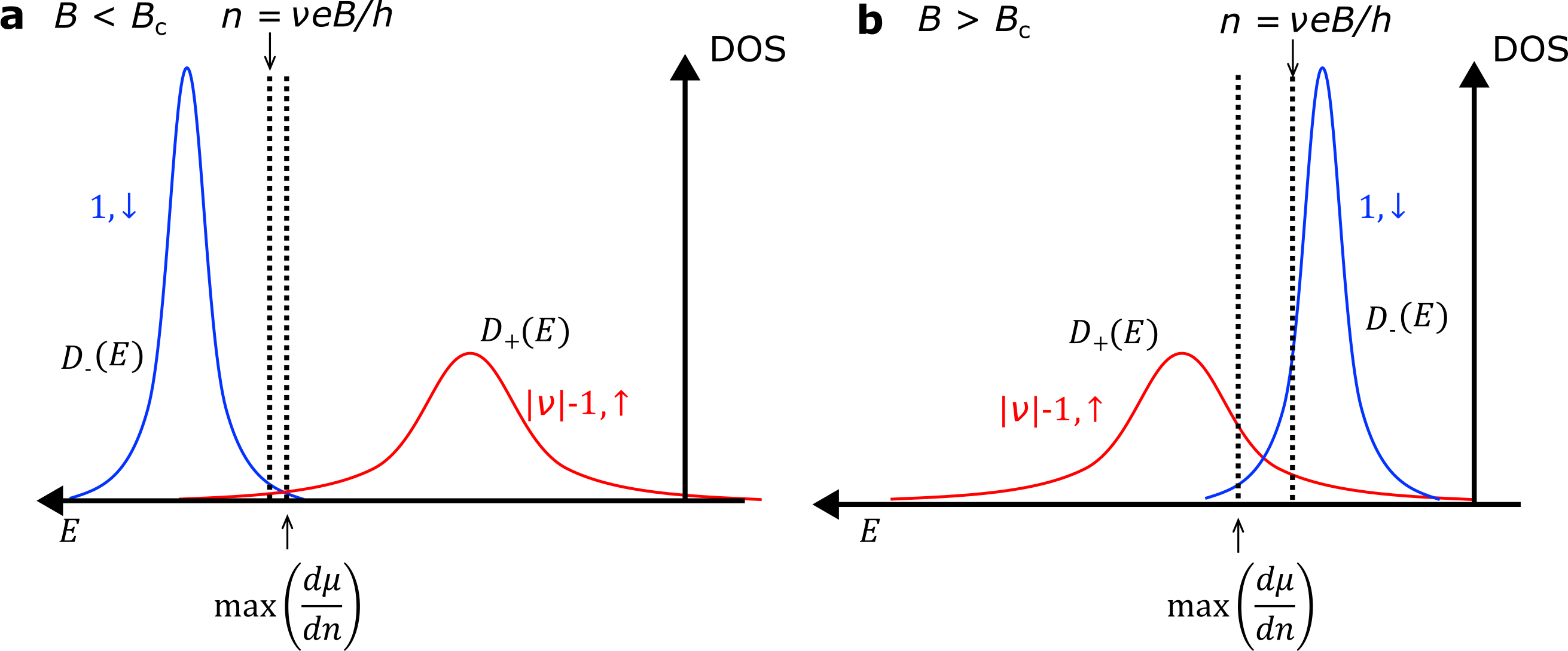}
    \caption{\textbf{Cartoon of disorder-broadened LLs near a phase transition.} \textbf{a}, Schematic of the density of states $D(E)$ of disorder broadened LLs at magnetic fields $B<B_{\rm{C}}$ below the first-order phase transitions. Holes first fill the $(\nu-1)^{\rm{th}}$ majority spin LL; the peak in d$\mu$/d$n$ occurs at slightly lower hole density than the density $n=\nu eB/h$. \textbf{b}, Similar but at magnetic fields $B>B_{\rm{C}}$, for which the minority spin LL is filled first. In this case, the peak  in d$\mu$/d$n$ occurs at higher hole density than the density $n=\nu eB/h$, and the effect is magnified because the LL spacing (gap) is smaller.} 
    \label{fig:DisorderCartoon}
\end{figure*}

At a given LL transition, there are two relevant LLs, the $N = |\nu|$ LL of majority spin and the lowest ($N=1$) minority spin LL. In general, these LLs are not necessarily equally broadened by disorder. The amount of broadening as a function of LL index can depend on details of the type of disorder in the sample \cite{ando_theory_1974,pereira_correlated_2011}, as well as interactions. Here we assume that the majority spin LL $N = |\nu|$ is more spread than the minority spin $N = 1$ LL, because that is more consistent with the experimental data. In particular, we take a functional form for the density of states of each LL to be given by
\begin{equation}
    D_{\pm}(E) = \frac{1}{\sqrt{2\pi} \sigma_{\pm}}\exp(- (E-E_{\pm})^2 / 2\sigma_{\pm}^2)
\end{equation}
where $+(-)$ refers to the majority (minority) spin LL. The Gaussian form of broadening is unimportant here; any form of disorder broadening will have a qualitatively similar effect. At a magnetic field below the transition, $E_+ > E_-$, and holes fill the majority spin LL first, while above the transition, $E_- > E_+$. Experimentally, the gap that we probe is nominally $|E_+ - E_-|$, except that finite disorder ($\sigma_{\pm} > 0$) will reduce the gap size from this maximal value. Importantly, when the LL width is comparable to the nominal gap size, there will be some overlap in the density of states associated with each respective LL. In this case, there are two important energies: the energy at which minimal density of states occurs, which will coincide with the peak in d$\mu$/d$n$, and the energy at which $n = \nu e B/h$, at which the integer quantum Hall gap nominally occurs. These two energies are no longer identical in the presence of finite disorder and asymmetric broadening. As illustrated in Fig. \ref{fig:DisorderCartoon}, for $E_+ > E_-$ $(E_+ < E_-)$, the peak of d$\mu$/d$n$ will occur at lower (higher) hole density relative to integer $\nu$. This effect is magnified when the spacing between the LLs is small and there is greater overlap between adjacent LLs, as is the case immediately above the transition. In the experiment, as the field gets larger and the gap increases, the density shifts slowly back toward the appropriate integer $\nu$, because the LL overlap will decrease as the gap gets larger.

\begin{figure*}[h]
    \renewcommand{\thefigure}{S\arabic{figure}}
    \centering
    \includegraphics{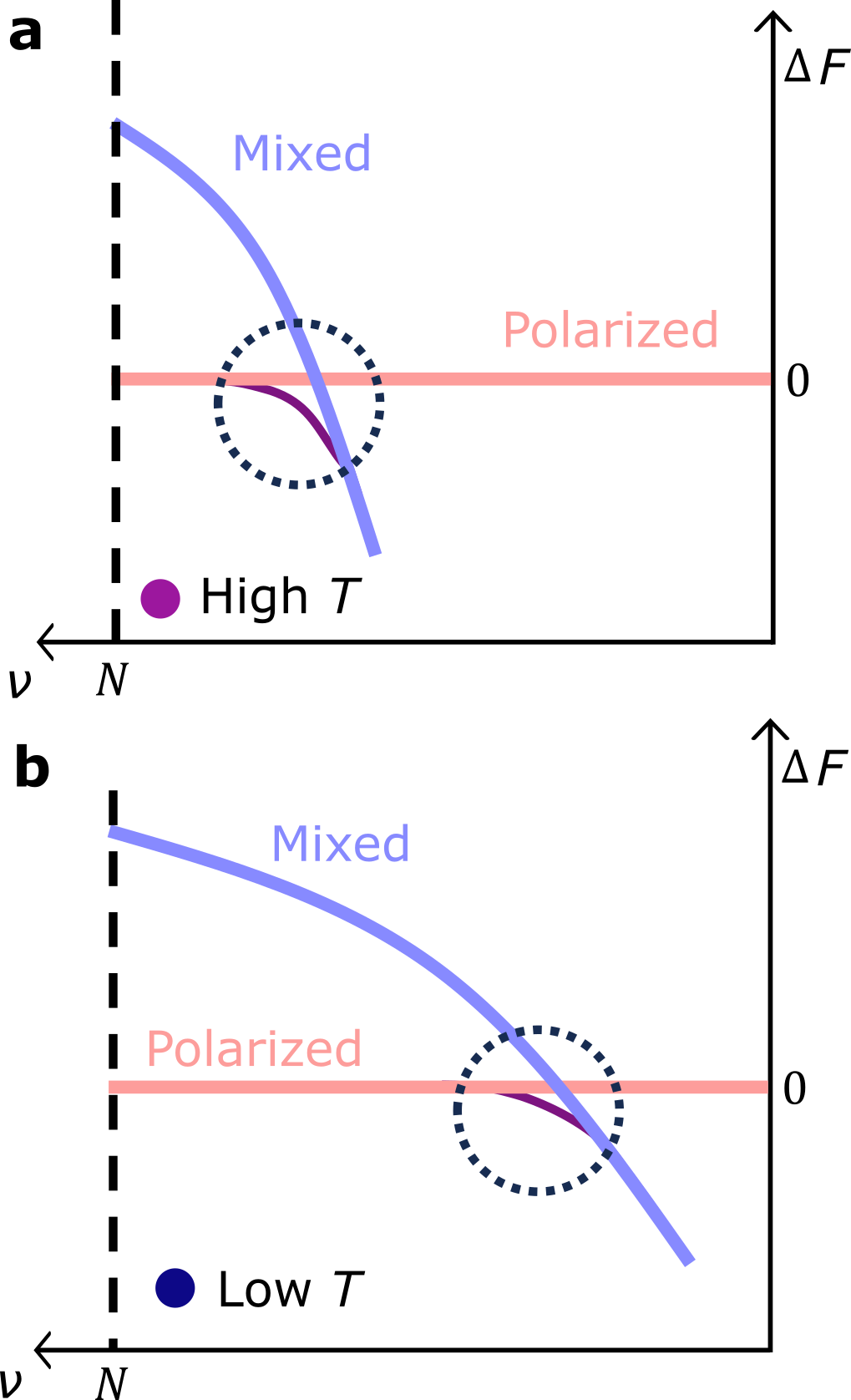}
    \caption{\textbf{Cartoon of free energy of mixed and polarized phases at various temperatures}. \textbf{a-b}, Schematic of the free energy $\Delta F = F_{\textrm{mixed}} - F_{\textrm{polarized}}$ of the mixed phase (the blue curve) relative to the polarized phase (the red straight line plotted as a reference) as a function of $\nu$ at high temperature (\textbf{a}) and low temperature (\textbf{b}). In the mixed phase, LLs of both spins are filled, while in the polarized phase only LLs of a single spin are filled. At low temperatures, the slope $\frac{\partial \Delta F}{\partial \nu}=\Delta \mu$ is less steep (see Eq.~\ref{eq:SommerfeldEq}) and the phase crossing point moves further away from the LL gap. Because the slope gets smaller at lower temperatures, the concavity of a line corresponding to a mixture of the two phases near the transition (purple) is smaller, leading to a weaker negative compressibility signal.
    }
    \label{fig:TempCartoon}
\end{figure*}

We next expand on the temperature dependence of the free energy of the two phases given the effects of disorder broadening of the LLs. For a fixed density $n$ and magnetic field $B$, the chemical potential $\mu$ at finite temperature $T$ can be approximated (via the Sommerfeld expansion) as 
\begin{equation}
    \mu(T,n) \approx \mu(T=0,n) - \xi \frac{\pi^2}{6}(k_BT)^2,\quad \xi \equiv \frac{\partial \log D(\epsilon)}{\partial \epsilon} \biggr|_{\epsilon=\mu(T=0,n)}
\end{equation}
where $D(\epsilon)$ is the density of states at energy $\epsilon$ and $k_B$ is Boltzmann's constant. From the schematic of Fig. \ref{fig:DisorderCartoon}, we can estimate $\xi$ at filling $n \gtrsim \nu eB/h$ (lower hole density than the LL gap), and we can observe that $\xi$ in the mixed regime ($\xi_{\rm{mixed}}$) is much larger than in the polarized state, so that $\Delta \xi = \xi_{\rm{mixed}} -\xi_{\rm{polarized}} > 0$. Thus, the difference in the chemical potentials between the two phases is approximated by 
\begin{equation}
    \Delta\mu(T,n) \approx \Delta\mu(T=0,n) - \Delta \xi\frac{\pi^2}{6}(k_B T)^2
    \label{eq:SommerfeldEq}
\end{equation}
In Fig. \ref{fig:TempCartoon} we show the difference in free energy $\Delta F$ between the mixed and polarized phases implied by this picture. The derivative $\frac{\partial \Delta F}{\partial n} = \Delta \mu$ is expressed in Eq.~\ref{eq:SommerfeldEq}. Because of the sign of $\Delta \xi$, $\Delta \mu$ will become more negative at higher temperatures, and equivalently the slope of $\Delta F$ will become steeper. This leads to both the crossing point (phase transition) occurring closer to the LL gap at higher temperatures and an overall greater concavity (corresponding to more negative $\frac{\partial \mu}{\partial n}$) at higher temperatures. At the highest temperatures, thermal broadening likely leads to a more weakly first order transition, so d$\mu$/d$n$ eventually weakens.

